\documentclass[twocolumn,twocolappendix]{aastex63} %,linenumbers

 \newcommand{\E}[1]{\ensuremath{\times 10^{#1}}}

\def\gtrsim{\mathrel{\hbox{\rlap{\hbox{\lower4pt\hbox{$\sim$}}}\hbox{$>$}}}}

\def \arcsec {\hbox{$^{\prime\prime}$}}

\usepackage{amsmath}%,siunitx}
\usepackage{enumitem}
\setlist[enumerate]{itemsep=-1.5mm}
\received{\today}%June 1, 2019
\revised{--}%January 10, 2019
\accepted{--}

\submitjournal{ApJS}

\shorttitle{DGPS Catalog}
\shortauthors{O'Connor et al.}
\graphicspath{{./}{figures/}}
\begin{document}

\title{The \textit{Swift} Deep Galactic Plane Survey (DGPS) Phase-I Catalog}

\correspondingauthor{Brendan O'Connor}
\email{oconnorb@gwmail.gwu.edu}

\author[0000-0002-9700-0036]{B. O'Connor}
    \affiliation{Department of Physics, The George Washington University, Washington, DC 20052, USA}
    \affiliation{Department of Astronomy, University of Maryland, College Park, MD 20742-4111, USA}
    \affiliation{Astrophysics Science Division, NASA Goddard Space Flight Center, 8800 Greenbelt Rd, Greenbelt, MD 20771, USA}
\author[0000-0003-1443-593X]{C. Kouveliotou}
    \affiliation{Department of Physics, The George Washington University, Washington, DC 20052, USA}
\author[0000-0002-8465-3353]{P.~A. Evans}
    \affiliation{School of Physics and Astronomy, University of Leicester,
University Road, Leicester, LE1 7RH, UK}
\author[0000-0002-1653-6411]{N. Gorgone}
    %\affiliation{Booz Allen Hamilton, 308 Sentinel Dr, Annapolis Junction, MD 20701, USA}
    \affiliation{Department of Physics, The George Washington University, Washington, DC 20052, USA}
\author[0000-0002-3905-4853]{A.~J. van Kooten}
    \affiliation{Department of Physics, The George Washington University, Washington, DC 20052, USA}
\author[0000-0003-0902-1935]{S. Gagnon}
    \affiliation{Department of Physics, The George Washington University, Washington, DC 20052, USA}
\author[0000-0002-8832-6077]{H. Yang}
    \affiliation{Department of Physics, The George Washington University, Washington, DC 20052, USA}
%%%%%%%%%%%%%%%%%%%%%%%%%%%%%%%%%%%%%
%%%%%%%Alphabetical below here%%%%%%%
%%%%%%%%%%%%%%%%%%%%%%%%%%%%%%%%%%%%%
\author[0000-0003-4433-1365]{M.~G. Baring}
    \affiliation{Department of Physics and Astronomy - MS 108, Rice University, 6100 Main Street, Houston, Texas 77251-1892, USA}
\author[0000-0001-8018-5348]{E. Bellm} 
    \affiliation{DIRAC Institute, Department of Astronomy, University of Washington, 3910 15th Avenue NE, Seattle, WA 98195, USA}
\author[0000-0001-7833-1043]{P. Beniamini}
    \affiliation{Department of Natural Sciences, The Open University of Israel, P.O Box 808, Ra'anana 4353701, Israel}
    \affiliation{Astrophysics Research Center of the Open university (ARCO), The Open University of Israel, P.O Box 808, Ra’anana 4353701, Israel}
    \affiliation{Department of Physics, The George Washington University, Washington, DC 20052, USA}
\author[0000-0003-0030-7566]{J. Brink}
    \affiliation{Department of Astronomy, University of Cape Town, Private Bag X3, Rondebosch 7701, South Africa}
    \affiliation{South African Astronomical Observatory, P.O. Box 9, Observatory 7935, Cape Town, South Africa}
\author[0000-0002-7004-9956]{D.~A.~H. Buckley}
    \affiliation{South African Astronomical Observatory, P.O. Box 9, Observatory 7935, Cape Town, South Africa}
    \affiliation{Southern African Large Telescope, P.O. Box 9, Observatory 7935, Cape Town, South Africa}
    \affiliation{Department of Astronomy, University of Cape Town, Private Bag X3, Rondebosch 7701, South Africa}
    \affiliation{Department of Physics, University of the Free State, P.O. Box 339, Bloemfonein 9300, South Africa}
\author[0000-0003-1673-970X]{S.~B. Cenko}
    \affiliation{Astrophysics Science Division, NASA Goddard Space Flight Center, 8800 Greenbelt Rd, Greenbelt, MD 20771, USA}
    \affiliation{Joint Space-Science Institute, University of Maryland, College Park, MD 20742 USA}
\author[0009-0001-0232-3968]{O.~D. Egbo}
    \affiliation{Department of Astronomy, University of Cape Town, Private Bag X3, Rondebosch 7701, South Africa}
    \affiliation{South African Astronomical Observatory, P.O. Box 9, 7935 Observatory, South Africa}
\author[0000-0002-5274-6790]{E. G\"{o}\u{g}\"{u}\c{s}}
    \affiliation{Sabanc\i~University, Faculty of Engineering and Natural Sciences, \.Istanbul 34956 Turkey}
\author[0000-0001-8530-8941]{J. Granot}
    \affiliation{Department of Natural Sciences, The Open University of Israel, P.O Box 808, Ra'anana 4353701, Israel}
    \affiliation{Astrophysics Research Center of the Open university (ARCO), The Open University of Israel, P.O Box 808, Ra’anana 4353701, Israel}
    \affiliation{Department of Physics, The George Washington University, Washington, DC 20052, USA}
\author{C. Hailey}
    \affiliation{Columbia Astrophysics Laboratory, Columbia University, New York, NY 10027, USA}
\author[0000-0002-8548-482X]{J. Hare}
    \affiliation{Astrophysics Science Division, NASA Goddard Space Flight Center, 8800 Greenbelt Rd, Greenbelt, MD 20771, USA}
     \affiliation{Center for Research and Exploration in Space Science and Technology, NASA/GSFC, Greenbelt, Maryland 20771, USA}
     \affiliation{The Catholic University of America, 620 Michigan Ave., N.E. Washington, DC 20064, USA}
\author[0000-0003-2992-8024]{F. Harrison}
    \affiliation{Cahill Center for Astrophysics, California Institute of Technology, 1216 East California Boulevard, Pasadena, CA 91125, USA}
\author[0000-0002-8028-0991]{D. Hartmann}
    \affiliation{Department of Physics and Astronomy, Clemson University, Kinard Lab of Physics, Clemson, SC 29634-0978, USA} 
\author[0000-0001-9149-6707]{A.~J. van der Horst}
    \affiliation{Department of Physics, The George Washington University, Washington, DC 20052, USA}
\author[0000-0002-1169-7486]{D. Huppenkothen}
    \affiliation{SRON Netherlands Institute for Space Research, Niels Bohrweg 4, 2333CA Leiden, The Netherlands}
\author{L. Kaper}
    \affiliation{University of Amsterdam, Science Park 904, 1098 XH Amsterdam, The Netherlands}
\author[0000-0002-6447-4251]{O. Kargaltsev}
    \affiliation{Department of Physics, The George Washington University, Washington, DC 20052, USA}
\author[0000-0002-6745-4790]{J.~A. Kennea}
    \affiliation{Department of Astronomy and Astrophysics, The Pennsylvania State University, 525 Davey Lab, University Park, PA 16802, USA}
\author[0000-0002-8286-8094]{K. Mukai}
    \affiliation{CRESST II and X-ray Astrophysics Laboratory, NASA/GSFC, Greenbelt, MD 20771, USA}
    \affiliation{Department of Physics, University of Maryland Baltimore County, 1000 Hilltop Circle, Baltimore MD 21250, USA}
\author[0000-0002-6986-6756]{P.~O. Slane}
    \affiliation{Center for Astrophysics, Harvard \& Smithsonian, 60 Garden St. Cambridge, MA 02138, USA}
\author[0000-0003-2686-9241]{D. Stern}
    \affiliation{Jet Propulsion Laboratory, California Institute of Technology, 4800 Oak Grove Drive, Mail Stop 169-221, Pasadena, CA 91109, USA}
\author[0000-0002-1869-7817]{E. Troja}
    \affiliation{University of Rome Tor Vergata, Department of Physics, via della Ricerca Scientifica 1, 00100, Rome, IT}
\author[0000-0003-2714-0487]{Z. Wadiasingh}
    \affiliation{Department of Astronomy, University of Maryland, College Park, Maryland 20742, USA}
    \affiliation{Astrophysics Science Division, NASA Goddard Space Flight Center, 8800 Greenbelt Rd, Greenbelt, MD 20771, USA}
     \affiliation{Center for Research and Exploration in Space Science and Technology, NASA/GSFC, Greenbelt, Maryland 20771, USA}
\author[0000-0002-3101-1808]{R.~A.~M.~J. Wijers}
    \affiliation{Anton Pannekoek Institute, University of Amsterdam, Postbus 94249, 1090 GE Amsterdam, The Netherlands}
    \affiliation{Department of Physics, The George Washington University, Washington, DC 20052, USA}
\author[0000-0002-6896-1655]{P. Woudt}
    \affiliation{Department of Astronomy, University of Cape Town, Private Bag X3, Rondebosch 7701, South Africa}
\author[0000-0002-7991-028X]{G. Younes}
    \affiliation{Astrophysics Science Division, NASA Goddard Space Flight Center, 8800 Greenbelt Rd, Greenbelt, MD 20771, USA}
    \affiliation{Department of Physics, The George Washington University, Washington, DC 20052, USA}

%%%%%%%%%%%%%%%%%%%%%%%%%%%%%%%%%%%%%%%%%%
%%%%%%%%%%%%%%%%%%%%%%%%%%%%%%%%%%%%%%%%%%

%%%%%%%%%%%%%%%%%%%%%%%%%%%%%%%%%%%%%%%%%%
%%%%%%%%%%%%%%%%%%%%%%%%%%%%%%%%%%%%%%%%%%



\begin{abstract}
The \textit{Swift} Deep Galactic Plane Survey is a \textit{Swift} Key Project consisting of 380 tiled pointings covering $\sim$40 deg$^{2}$ of the Galactic Plane between  longitude $10$\,$<$\,$|l|$\,$<$\,$30$ deg and latitude $|b|$\,$<$\,$0.5$ deg.  Each pointing has a $5$ ks exposure, yielding a total of 1.9 Ms spread across the entire survey footprint. Phase-I observations were carried out between March 2017 and May 2021. The Survey is complete to depth $L_X$\,$>$\,$10^{34}$ erg s$^{-1}$ to the edge of the Galaxy. The main Survey goal is to produce a rich sample of new X-ray sources and transients, while also covering a broad discovery space. Here, we introduce the Survey strategy and present a catalog of sources detected during Phase-I observations. In total, we identify 928 X-ray sources, of which 348 are unique to our X-ray catalog. We report on the characteristics of sources in our catalog and highlight sources newly classified and published by the DGPS team. 
\end{abstract}



\section{Introduction}

Since the inception and discovery of X-ray astronomy, from the detection of Sco X-1 and the launch of the first X-ray satellite in 1970 \citep[\textit{Uhuru};][]{uhuru}, a diverse assortment of X-ray emitting sources have been discovered and sorted into numerous distinct classes. These classes include chromospheric activity from young stars, cataclysmic variables (CVs), symbiotic binaries, young stellar objects (YSOs),  magnetars, and X-ray binaries comprising  a compact object, either a neutron star (NS) or black hole (BH), and a low-mass (LMXBs) or high-mass (HMXBs) star. 

Within our Galaxy, the brightest X-ray sources are known to be X-ray binaries with peak X-ray luminosities in excess of $L_X$\,$>$\,$10^{36-39}$ erg s$^{-1}$. However, our Milky Way (MW) also hosts a significant population of faint X-ray sources ($L_X$\,$<$\,$10^{33-35}$ erg s$^{-1}$) \citep{Muno2005a,Muno2005b,Degenaar2009,Degenaar2010}. These sources are likely dominated by magnetic CVs \citep{Barrett1999,Wang2002,Revnivtsev2009,Pretorius2013}, quiescent LMXBs \citep{Muno2005a,Muno2005b}, and quiescent magnetars \citep{CotiZelati2018}, among others. Their discovery is crucial to expand our understanding of their source populations and their formation pathways within our Galaxy. 
%quiescent magnetars, polar CVs etc...

X-ray surveys of the Galactic Plane (GP) present a prime opportunity for discovery of these faint sources. Thus far, sensitive and high-resolution X-ray satellites, such as \textit{XMM-Newton} or \textit{Chandra} \citep{Wijnands2006,Jonker2011,xmmsurvey2013}, have been used to search for serendipitous faint X-ray sources within the true target's field of view. Such procedures, however, are not uniform in depth nor do they cover the full extent of the GP, relying instead on pointings directed at known bright sources. 
Therefore, dedicated, homogeneous X-ray surveys are required to identify the population and number of faint X-ray sources within the Galaxy. 

The \textit{Neil Gehrels Swift Observatory} \citep{Gehrels2004} X-ray Telescope \citep[XRT;][]{Burrows2005} utilizes a CCD detector with sensitivity to X-ray photons over the range $0.3$\,$-$\,$10$ keV. The instrument field of view (FOV) is $23.6\arcmin\times 23.6\arcmin$ with an effective area of 110 cm$^2$ at 1.5 keV and an angular resolution of $18\arcsec$. The low background ($10^{-6}$ cts s$^{-1}$ pix$^{-1}$; \citealt{Evans2014}), arcsecond source localization, and fast slew rate, make the \textit{Swift}/XRT optimal for surveys of crowded environments, such as the GP \citep{Reynolds2013}, Small Magellanic Cloud \citep{Kennea2018}, and the Galactic Bulge \citep{Shaw2020,Bahramian2021}. 

Here, we outline our \textit{Swift} Deep Galactic Plane Survey (DGPS) strategy and present the catalog of sources detected in Phase-I observations across the $\sim$40 deg$^2$ portion of the GP covered by the DGPS. We present the survey design and strategy in \S \ref{sec: footprint}. In \S \ref{sec: obs/analysis} we discuss our source detection procedures and the process for creating a unique source catalog. The catalog results, discussion of implications, and overall conclusions are presented in \S \ref{sec: results}, \S \ref{sec: discussion}, and \S \ref{sec: conclusions}, respectively.

\begin{figure}[h]
\centering
\includegraphics[width=\columnwidth]{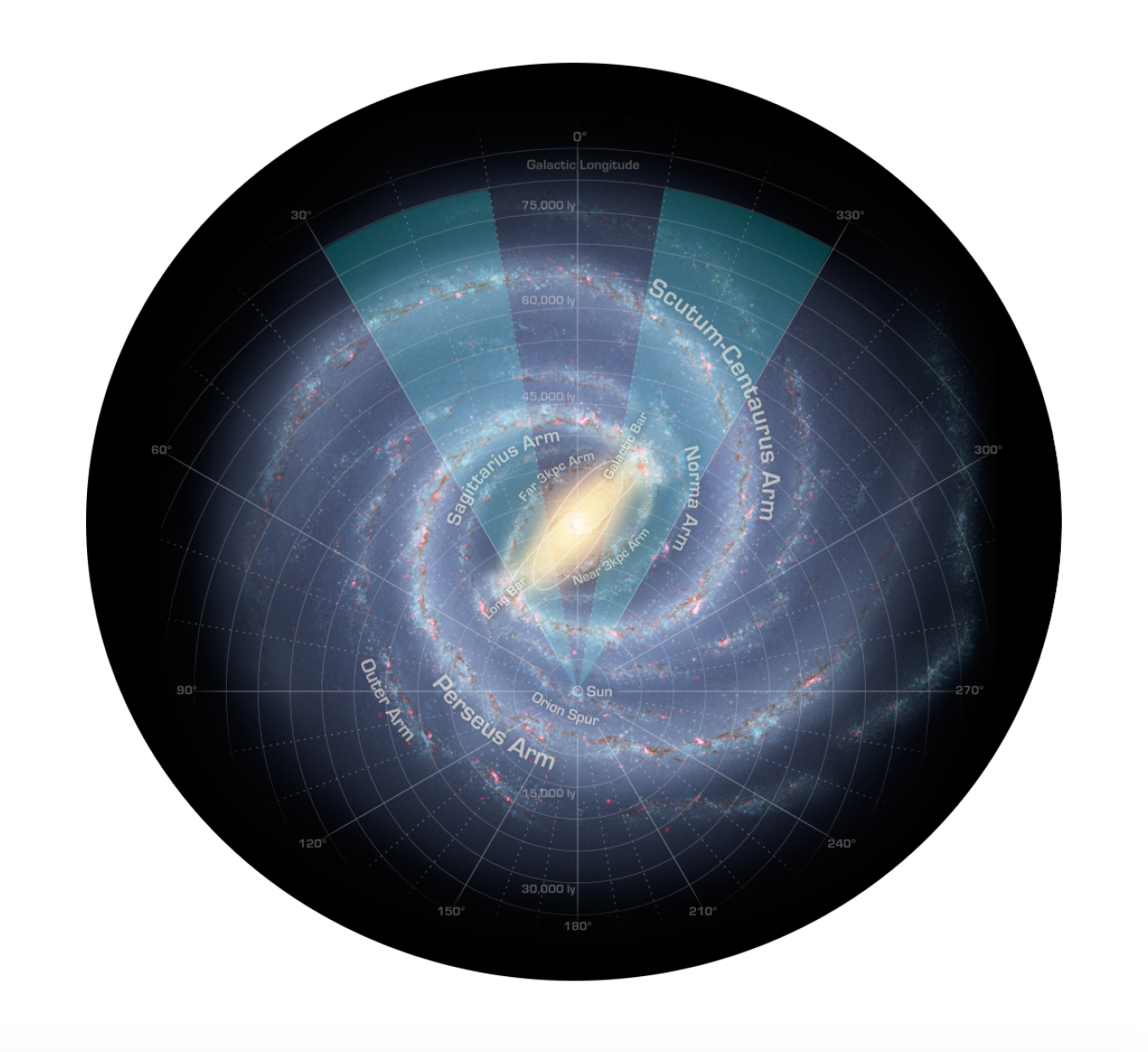}
\caption{The shaded blue regions show the line of sight of the DGPS survey footprint through the Milky Way's disk. The background image is an illustration of the Milky Way with credit to NASA/JPLCaltech/ESO/R. Hurt.} 
\label{fig: MWfig}
\end{figure}

\section{Survey Footprint and Observing Strategy}
\label{sec: footprint}

%\textcolor{blue}{Read this paper to get ideas https://arxiv.org/pdf/1012.1469.pdf and https://www.aanda.org/articles/aa/pdf/2010/15/aa13570-09.pdf and https://ui.adsabs.harvard.edu/abs/2004MNRAS.351...31H/abstract}

%CONFIRMED: 380 unique pointings but 8 were duplicated so 388 unique target ID were used! No need to mention duplicates specifically in the paper.

\begin{figure*} 
\centering
\includegraphics[width=2.15\columnwidth]{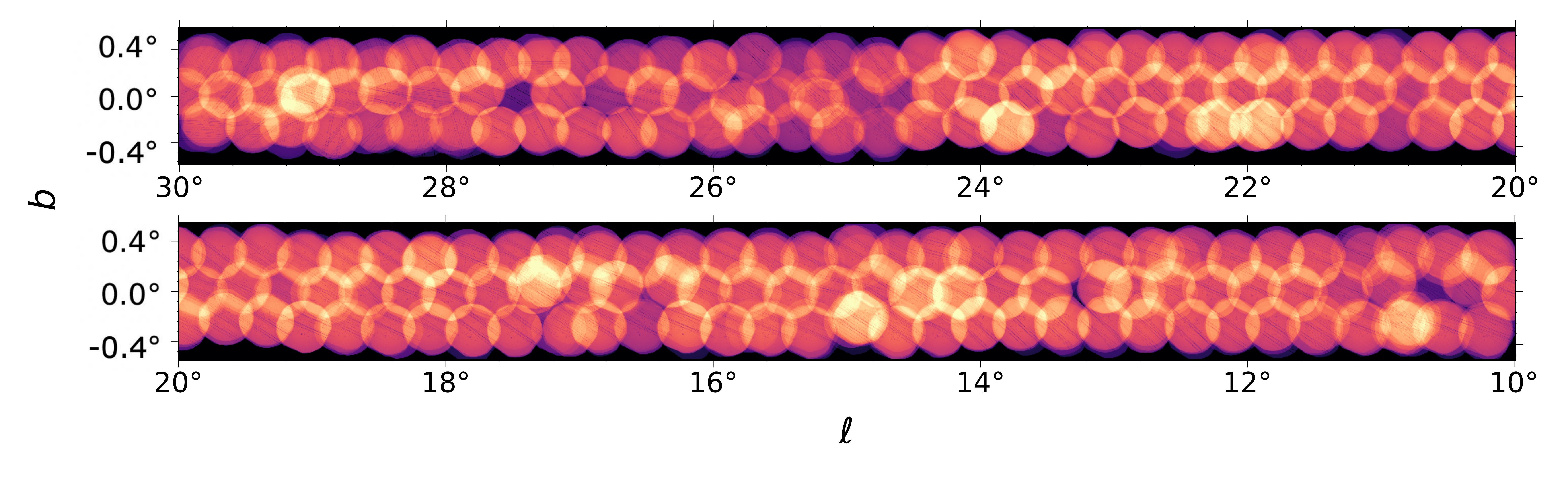}
\includegraphics[width=2.15\columnwidth]{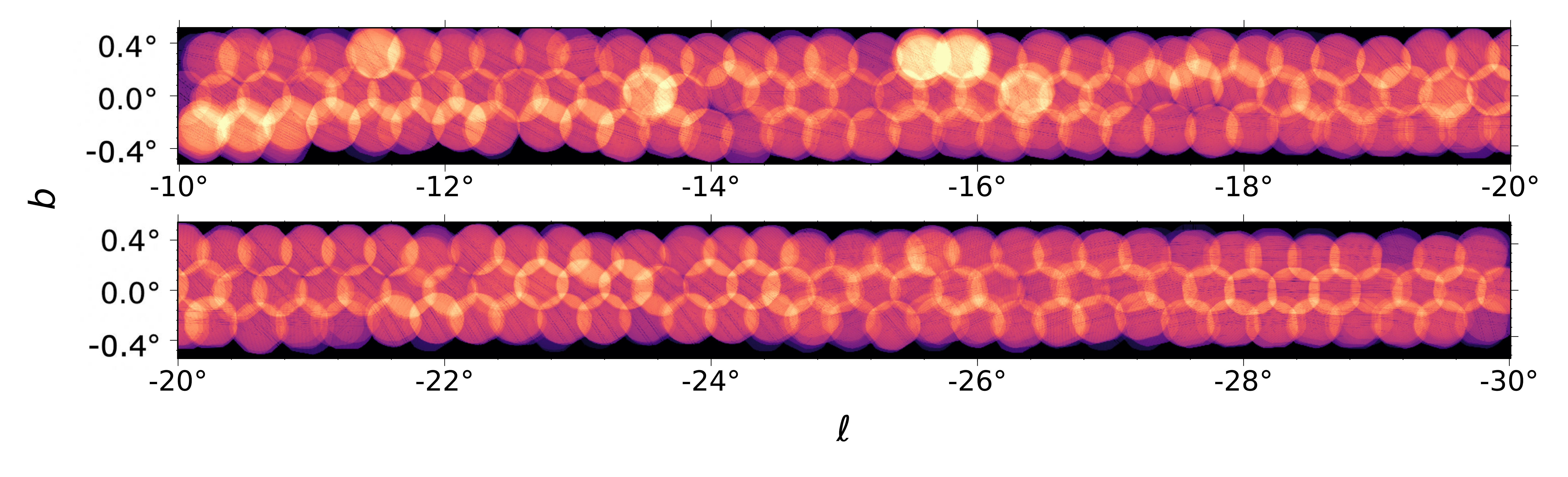}
\caption{\textit{Swift}/XRT exposure map of the DGPS footprint. The $4\arcmin$ overlap region between adjacent tiles is clearly demonstrated. A few tile positions were serendipitously observed twice, leading to a higher exposure (brighter regions). The median exposure across all pixels is 4.6 ks. % with a maximum exposure of 21.3 ks. %The interquartile range is between 3.8 and 6.4 ks, with 90\% of values below 7.8 ks. 
The variation in exposure in the two observed regions of the GP is negligible.
}
\label{fig: expmap_mosaic}
\end{figure*}

\begin{figure*} 
\centering
\includegraphics[width=2.15\columnwidth]{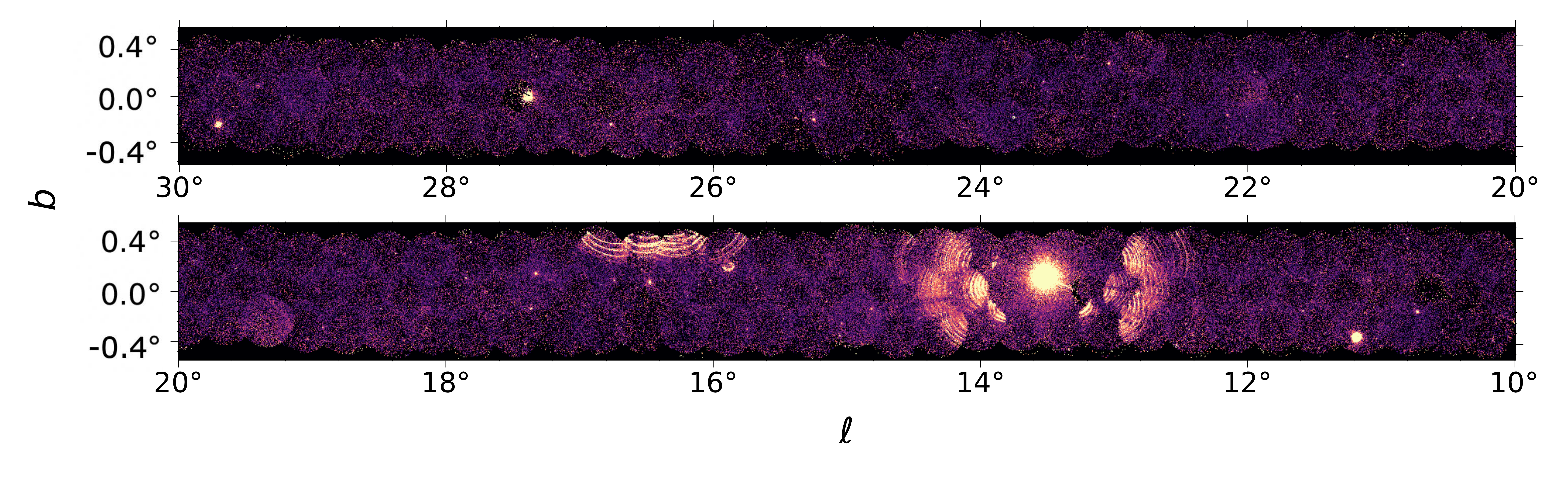}
\includegraphics[width=2.15\columnwidth]{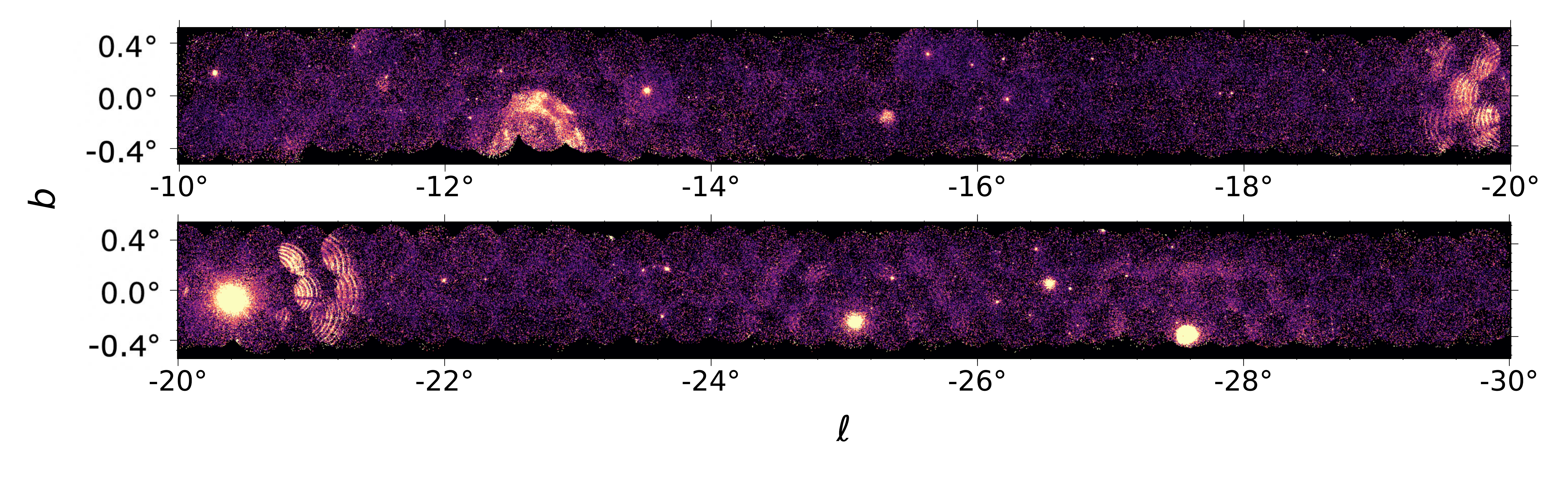}
\caption{Full XRT band ($0.3$\,$-$\,$10$ keV) mosaic of the Galactic plane using an Aitoff projection in Galactic coordinates. The image covers the full footprint of DGPS Phase-I. The pixel size is $4.7\arcsec$/pix, and images have been smoothed with a Gaussian kernel (with FWHM of 3 pixels) to improve visual clarity. The image has been divided by the exposure map (Figure \ref{sec: footprint}) in order to smooth out exposure related background in the overlapping regions.  The dominant sources of stray light at $l$\,$\approx$\,$-25^\circ$, $-20.5^\circ$, and $13.5^\circ$ are the LMXBs 4U 1624-49, 4U 1642-45, and GX 13+01. 
}
\label{fig: GP_mosaic}
\end{figure*}

The \textit{Swift} Deep Galactic Plane Survey (PI: C. Kouveliotou) is a \textit{Swift} Key Project and \textit{NuSTAR} Legacy Program\footnote{\url{https://www.nustar.caltech.edu/page/59\#g9}} covering $\sim$40 deg$^2$ of the GP (Figure \ref{fig: MWfig}) between Galactic longitude  $10$\,$<$\,$|l|$\,$<$\,$30$ deg and latitude $|b|$\,$<$\,$0.5$ deg. The total sky coverage of the Survey is 36 deg$^2$ when correcting for tile overlaps and the shape of the XRT FOV.  
The Survey encompasses 380 unique XRT pointings (see Figures \ref{fig: expmap_mosaic} and \ref{fig: GP_mosaic}), each observed for $\sim$\,$5$ ks for a total of $\sim$\,$1.93$ Ms exposure carried out between March 2017 to May 2021. Approximately half of these observations were performed between 2017 and 2019, and the second half between 2020 and 2021. All observations were performed with the \textit{Swift} X-ray Telescope \citep[XRT;][]{Burrows2005} in Photon Counting (PC) mode.

The design of our survey (latitude and longitude range; Figures \ref{fig: expmap_mosaic} and \ref{fig: GP_mosaic}) was driven by our primary science goal of thoroughly characterizing the magnetar and HMXB populations in the MW by their persistent emission, while avoiding the crowded Galactic center (Figure \ref{fig: MWfig}). We have additionally selected the Survey footprint such that each tile has a $4\arcmin$ overlap with its neighbor, taking into account the $23.6\arcmin$ XRT FOV.%, which allows for small portions of our survey reaching $\sim15-20$ ks exposure. These portions are therefore more sensitive to very faint X-ray sources.

\begin{figure} 
\centering
\includegraphics[width=\columnwidth]{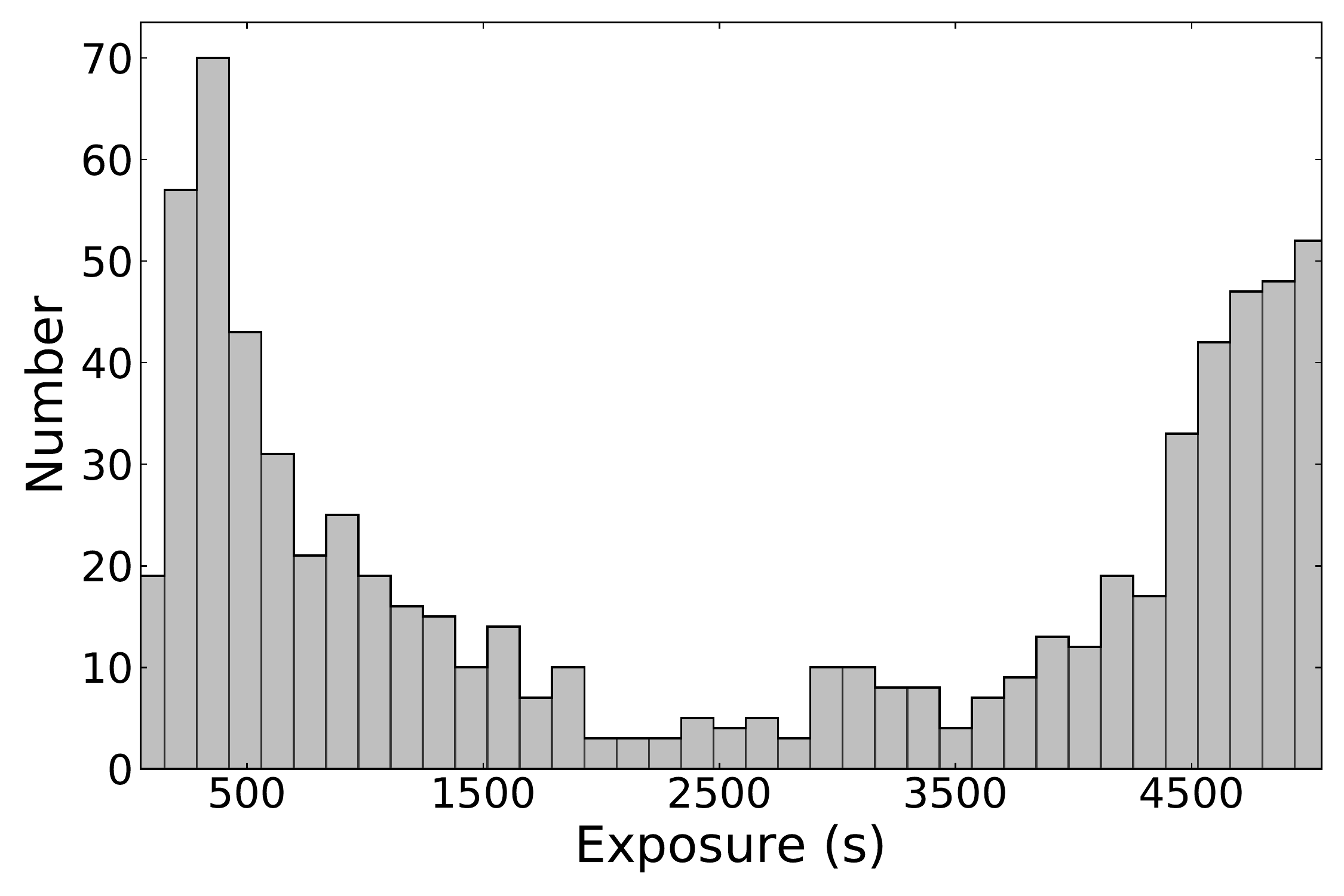}
\includegraphics[width=\columnwidth]{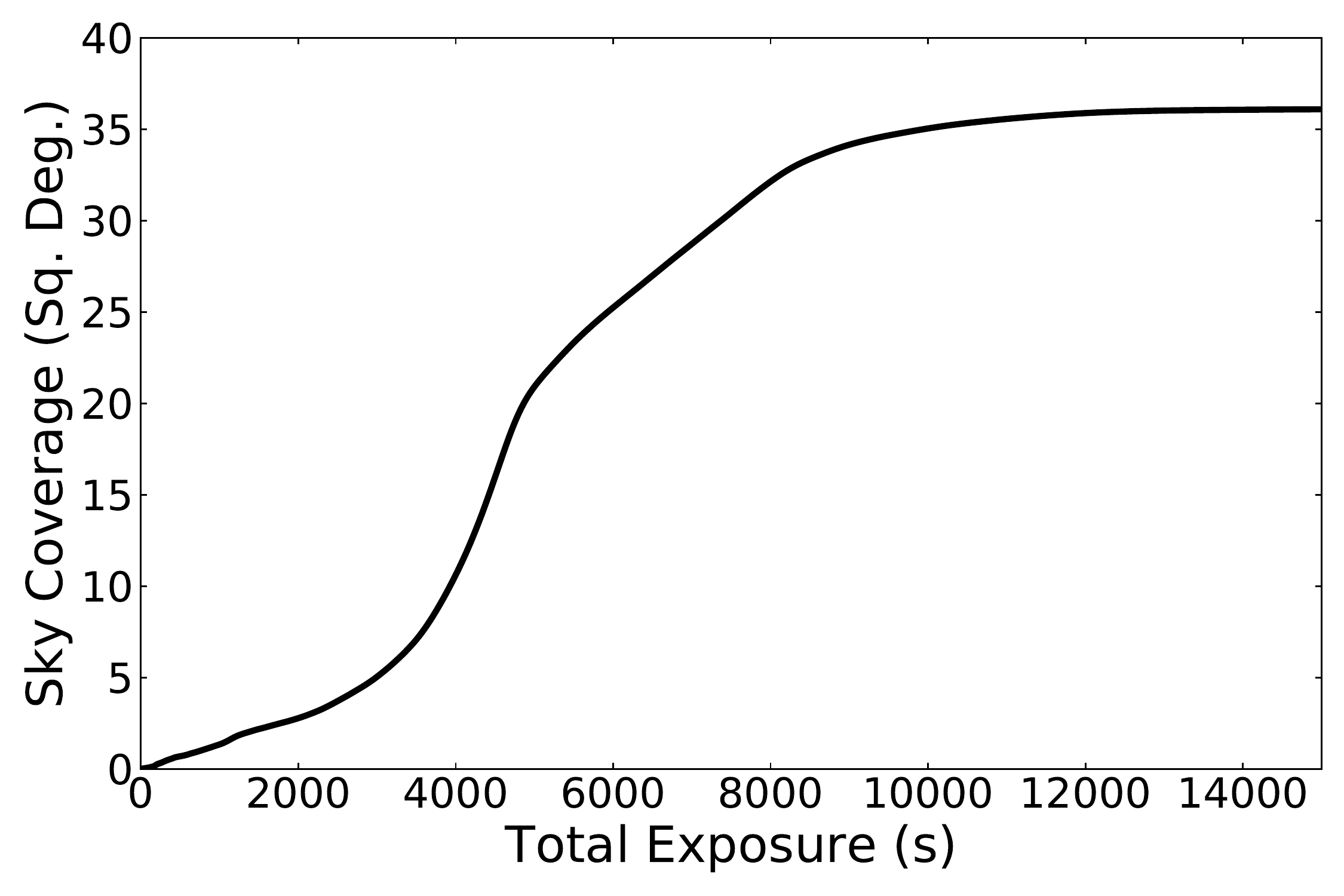}
\caption{\textbf{Top:} Histogram of single-epoch \textit{Swift}/XRT exposure time for all DGPS observations. 
\textbf{Bottom:} Sky coverage as a cumulative function of the exposure time (corrected for vignetting and bad pixels) across the entire DGPS Survey footprint after mosaicing all observations. The total sky coverage of the Survey is 36 deg$^2$. 
The median exposure time is 4.6 ks. 
The overlap regions between tiles lead to a higher exposure of up to 15-20 ks. 
}
\label{fig: obs_exp}
\end{figure}

In total, the Survey comprises 769 
%754+8 (pilot survey) = 762
%The 8 pilot survey tiles are 34689-34697 
observations\footnote{An observation is defined as all exposures covering a specific pointing obtained within a single UT day.} with \textit{Swift} covering the 380 pointings (Figure \ref{fig: expmap_mosaic}), including those observed during the DGPS Pilot Survey. 
This is due to the fact that in most cases ($\sim$70\%) multiple observations of the same field were required to yield a total of 5 ks exposure. 
In Figure \ref{fig: obs_exp}, we display a histogram of exposure times for these 769 %754+8 (pilot survey) = 762
single-epoch observations. 
We note that although a significant fraction (47\%) of single-epoch observations consisted of less than 2 ks of exposure, the median cumulative exposure across the Survey footprint is 4.6 ks (Figures \ref{fig: expmap_mosaic}  and \ref{fig: obs_exp}).
The fact that many tiles were observed multiple times was extremely useful for the identification of variable X-ray sources (see \S \ref{sec: variable} and \S \ref{sec: newly classified}).

On average, the survey is complete (\S \ref{sec: completeness}) to a depth of $L_X$\,$>$\,$1.0\times 10^{34}$ erg s$^{-1}$, to the edge of the Galaxy. However, it affords source detection to limits of $L_X$\,$\sim$\,$1.0\times 10^{33}$ erg s$^{-1}$ out to $\sim$\,$3$\,$-$\,$6$ kpc.

\section{\textit{Swift}/XRT Data Analysis}
\label{sec: obs/analysis}

Here, we outline our process for analyzing all 769 DGPS observations. Due to the long-term nature of the project, and the need for XRT to return to the same field multiple times (Figure \ref{fig: obs_exp}), we performed an initial analysis of all data when it was first obtained (\S \ref{sec: quicklook}). After the end of Phase-I observations, we performed a final processing (\S \ref{sec: finalcat}) of all observations to create the DGPS Phase-I catalog.

To do this, we performed source detection on mosaics of the DGPS observations (\S \ref{tab: main_cat}). Following the creation of a unique source catalog, we pulled additional information (e.g., flux, hardness ratio (HR), variability) from the Living Swift-XRT Point-source catalog\footnote{\url{https://www.swift.ac.uk/LSXPS/docs.php}} (LSXPS; \citealt{LSXPS}). LSXPS has processed all \textit{Swift}/XRT observations, including those comprising the DGPS, and this step avoids redundancy in re-processing all of the data and increases the overall scientific impact by allowing us to have an improved grasp on the source characteristics.

Through this process, we discovered that there exists a subset of DGPS sources ($\sim14\%$) that are not in the LSXPS catalog (\S \ref{sec: nonlsxps}). These sources lack some of the additional information that comes from LSXPS (e.g., HR variability), and we discuss their significance further in \S \ref{sec: nonlsxps} and \S \ref{sec: non-lsxps_match}. 
All sources detected by DGPS in our analysis of the mosaics (e.g., Figure \ref{fig: mosaic_ex}), including those not found in LSXPS,  are incorporated into our full catalog (see also Appendix \ref{sec: catalog properties}), and this includes those sources which were not identified in LSXPS processing \citep{LSXPS}.

\subsection{Quick-look Analysis}
\label{sec: quicklook}

The identification and prompt follow-up of variable or transient sources detected as part of the Survey required a rapid analysis of quick-look data\footnote{\url{https://www.swift.ac.uk/archive/ql.php}} as these became available $\sim$\,$2-6$ hours after the observations. %\footnote{Quick-look data is typically available to $\sim$\,$2-6$ hours after the observations.}. 
Quick-look data are not the final fully processed data, and are instead treated as a preliminary first look in order to identify sources displaying variability on a shorter timescale than the fully processed data are available ($\sim$\,$1$\,$-$\,$2$ weeks after the quick-look data\footnote{\url{https://swift.gsfc.nasa.gov/quick-look/swift\_process\_overview.html}}). The former data, however, allowed for rapid multi-wavelength follow-up observations. The single-epoch quick-look data was initially processed within a day of each XRT observation. 

In many cases ($\sim$\,$70\%$), \textit{Swift} did not perform the full $\sim$\,$5$ ks exposure in a single epoch (see Figure \ref{fig: obs_exp}). Therefore, in order to reach the full exposure for each tile, \textit{Swift} carried out multiple observations\footnote{In most cases, it took \textit{Swift} three observations of varying length for an individual tile to reach $5$ ks exposure.}, sometimes taken months apart. We utilized this to better identify variability by comparing the source flux between each observation. %This method was the most successful at determining sources for multi-wavelength follow-up.
We additionally checked archival flux values from available X-ray catalogs. 
We selected previously unknown or unclassified variable sources with an unabsorbed X-ray flux brighter than $F_X$\,$>$\,$1.0\times10^{-12}$ erg cm$^{-2}$ s$^{-1}$ ($0.3$\,$-$\,$10$ keV) for Target of Opportunity (ToO) follow-up with a variety of X-ray satellites, such as \textit{XMM-Newton}, \textit{Chandra}, \textit{NuSTAR}, and \textit{NICER}, through our approved programs, see, e.g., \citet{Gorgone2019,Gorgone2021,OConnor2021,OConnor2023polar,OConnor2023IP}.  

%The final number of unique sources detected with our quick-look pipeline is 290, of which 90 sources satisfied $F_X$\,$>$\,$1.0\times10^{-12}$ erg cm$^{-2}$ s$^{-1}$. The majority of these sources were previously known, but those which were unclassified and displaying variable behavior were followed up through our ToO programs. This rapid multi-wavelength follow-up of variable or transient sources identified through our quick-look analysis has led to multiple publications (\S \ref{sec: newly classified}), including the source classification of a LMXB \citep{Gorgone2019}, a Be/X-ray Binary \citep{OConnor2021}, an intermediate polar (IP) CV \citep{Gorgone2021}, and a polar magnetic CV \citep{OConnor2023polar}. The classification of additional sources targeted through our follow-up campaign is ongoing. 

\subsection{Final Image Processing and Source Detection}
\label{sec: finalcat}

\begin{figure*} 
\centering
\includegraphics[width=1.8\columnwidth]{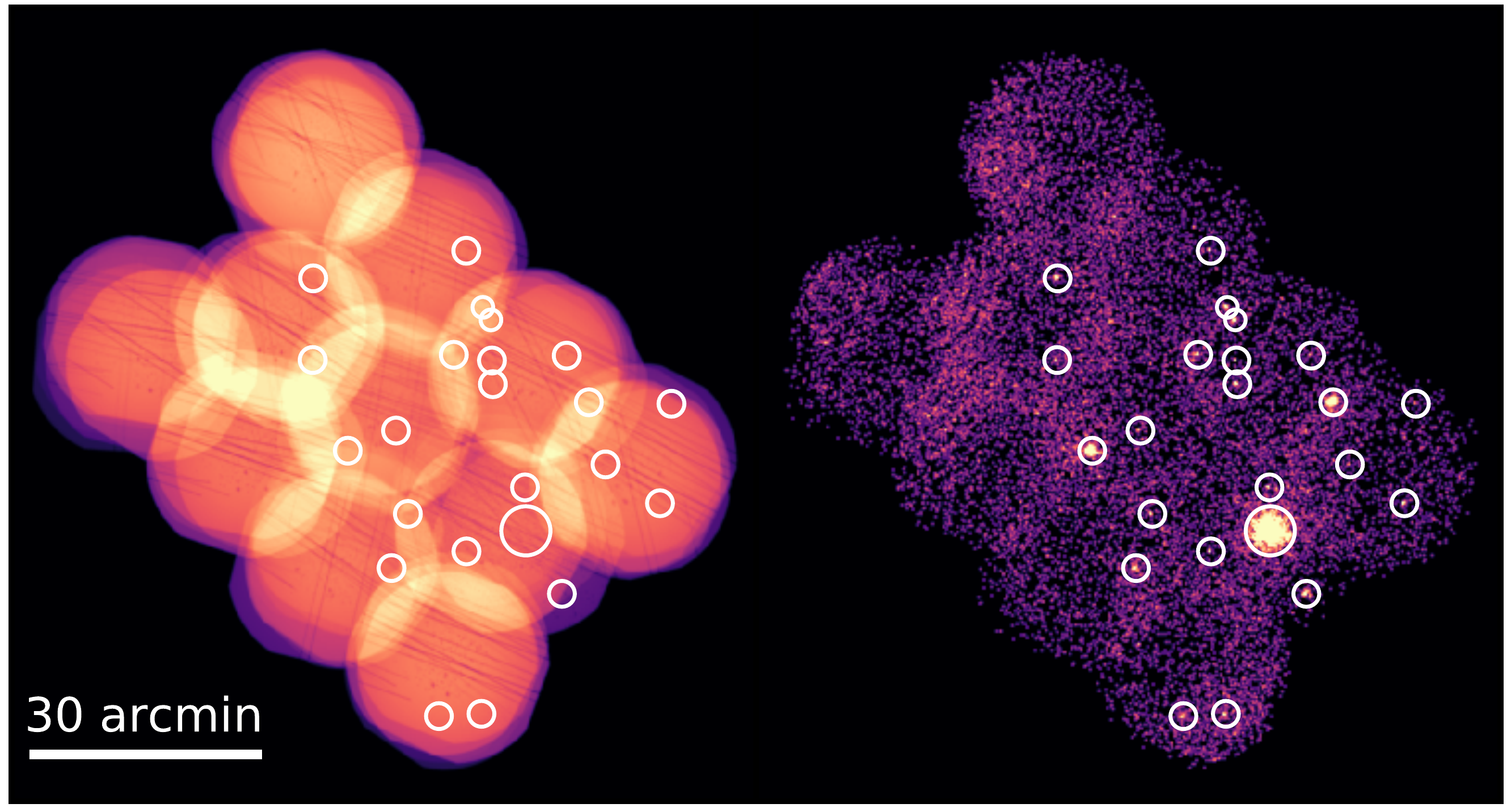}
\caption{Example exposure map and science image ($0.3$\,$-$\,$10$ keV) of a DGPS mosaic. The mosaic is centered at $l, b$ = 333.87$^\circ$, 0.026$^\circ$. White circles represent the location of sources detected in this image. 
The bright source to the right of the image is MAXI J1651-501, a Type-I X-ray Burster uncovered through DGPS observations \citep{Gorgone2019}. A weak stray light pattern (concentric bands) is visible on the left end of the mosaic. The images have been re-binned ($7.07\arcsec$/pix) and smoothed with a Gaussian kernel (with FWHM of 3 pixels) for display purposes.} 
\label{fig: mosaic_ex}
\end{figure*}

The rapid quick-look analysis of DGPS observations does not reach the full depth of the Survey. In order to produce a complete source catalog, we turned towards a more robust, yet computationally intensive, data analysis pipeline used to generate previous \textit{Swift} X-ray catalogs \citep{Evans2014,Evans2020}. This pipeline allows for the mosaicing of all observations within the DGPS. However, the \textit{Swift} DGPS covers $\sim$40 deg$^2$ of the Galactic Plane and performing source detection on regions of this size is not feasible due to the computational cost. Therefore, in order to reduce the computation time, while still achieving the maximum exposure across every part of the Survey, we defined 124 small mosaics covering the entire Phase-I Survey area. 
%What we did was take the b=0 tiles and sort through them taking all tiles within 23.6*1.4 = 33.04 arcmin of the central pointing of each b=0 tile. This was done for both sides of the plane, yielding 124. 
The mosaics were created such that there is an overlap for every mosaic, which means that some pointings were part of multiple mosaics. This ensures that every possible overlap of tiles was accounted for and allowed us to obtain the maximum exposure at every location within the DGPS footprint. An example mosaic is displayed in Figure \ref{fig: mosaic_ex}. 

The image processing, mosaic creation, and source detection algorithm are described in detail in \citet{Evans2014,Evans2020}. The pipeline made use of \texttt{HEASoft 6.29}. The iterative source detection procedure classifies each source using numerous quality flags, such as `\textit{good}', `\textit{reasonable}', or `\textit{poor}' (see \citet{Evans2014,Evans2020} for details\footnote{\url{https://www.swift.ac.uk/2SXPS/docs.php}}.) These flags indicate the level of significance of the detection and were calibrated using simulations of point sources. The false positive rate for \textit{good} sources is 0.3\%, and increases to 1\% when also including \textit{reasonable} sources, whereas including \textit{poor} sources yields a rate of spurious sources on the order of 10\% \citep{Evans2020}. These false positive rates are considered cumulative, and we note that the actual false positive rate for \textit{reasonable} and \textit{poor} sources is $\sim$\,$7\%$ and $\sim$\,$35\%$, respectively. Therefore, we remove sources with a \textit{poor} quality flag.

The \citet{Evans2020} pipeline also includes quality flags to prevent spurious sources in regions contaminated by stray light or extended sources (e.g., supernova remnants) as well as sources which are possible aliases of bright sources \citep[see Table 5 of][]{Evans2014}. We have excluded all sources occurring in the PSF of extremely bright sources, in regions of stray light or known extended objects, as well as those due to optical loading\footnote{\url{https://www.swift.ac.uk/analysis/xrt/optical_loading.php}}. The field flags were set manually by \citet{Evans2020}. 

After removing all sources with quality flags, we began by merging all blindly-detected sources in the same mosaic across the different energy bands. Source detection is run independently in four energy bands\footnote{The same energy bands were previously used in the 1SXPS, 2SXPS, and LSXPS catalogs \citep{Evans2014,Evans2020,LSXPS}.}%\footnote{These energy bands differ from those used in the original quick-look analysis. For the full catalog we adopted the same energy bands as previously used in the 1SXPS and 2SXPS catalogs \citep{Evans2014,Evans2020}, whereas the quick-look energy bands agreed with 1SWXRT \citep{DElia2013}.}
: the soft band (SB; $0.3$\,$-$\,$1$ keV), medium band (MB; $1$\,$-$\,$2$ keV), hard band (HB; $2$\,$-$\,$10$ keV), and the full band (FB; $0.3$\,$-$\,$10$ keV).  %We note that these energy bands are different from those used during our quick-look analysis. %this is already a footnote
We merged sources that were identified as the same source, but in different energy bands, by defining a match as either being within 10 pixels (1 pixel = 2.36$\arcsec$) or consistent at the 99.7\% level using Rayleigh statistics. At this stage, we include only the statistical position errors as each source within a single mosaic has the same astrometric solution. This process yields a list of unique sources identified in each mosaic.

As there is a one tile overlap between each mosaic, there are some duplicate sources that must be removed. We therefore cross-matched the source lists between every mosaic in order to remove duplicate sources which were consistent at the $99.7\%$ confidence level (including both the statistical and systematic error on the source positions). We are then left with a unique list of  %\textcolor{red}{802 (do we note the non-LSXPS in this number?)} 
sources detected across the entire DGPS footprint. 

The source count rates and fluxes in each energy band were then pulled from LSXPS%Living Swift-XRT Point-source catalog\footnote{\url{https://www.swift.ac.uk/LSXPS/docs.php}} (LSXPS; \citealt{LSXPS}) 
using the API tool\footnote{\url{https://www.swift.ac.uk/API/}}. We determined the LSXPS counterpart to each DGPS source using a radius of $20\arcsec$ or the 99.7\% combined error radius. As the LSXPS is a low-latency, continuously updated catalog, we note that our cross-match was performed on the LSXPS catalog of 2022 August 31. We note that we only include sources in LSXPS that are detected in our DGPS mosaics, and, therefore, only sources to the completeness limits of the DGPS.

The count rates were converted to a $0.3-10$ keV flux assuming a power-law spectrum with photon index $\Gamma$\,$=$\,$1.7$ and the Galactic hydrogen column density in the source direction from \citet{Willingale2013}. %\footnote{We clarify here that these methods for correcting the source flux are in contrast to our quick-look analysis. Instead, the main survey methods were chosen to agree with 1SXPS and 2SXPS.}. 
We further took from the LSXPS catalog the hardness ratios HR$_1$=M$-$S/M$+$S, HR$_2$=H$-$M/H$+$M, and the Pearson's $\chi^2$ probability that each source is variable based on their LSXPS lightcurves binned by observation. 

The source positions in LSXPS are based on either standard or astrometric positions. We therefore used the API tool to build XRT enhanced positions \citep{Goad2007,Evans2009} for all DGPS sources. We successfully built enhanced positions for 290 sources, and we accepted the position with the smallest error. We used the final source positions to name DGPS sources in the format: ``DGPS JHHMMSS.S$\pm$DDMMSS". 

All sources and their properties (along with LSXPS ID; \citealt{LSXPS}) are displayed in Table \ref{tab: main_cat}\footnote{Table \ref{tab: main_cat} is available in electronic form at the
CDS via anonymous ftp to cdsarc.u-strasbg.fr (130.79.125.5) or via \url{http://cdsweb.u-strasbg.fr/cgi-bin/qcat?J/ApJS/}.}. 
We detected a total of 802 sources of which 784 are detected in the FB, 724 in the HB, 668 in the MB, and 564 in the SB. %This is a factor of $\sim 3$ more sources than identified through the quick-look analysis of single-epoch observations (see \S \ref{sec: quicklook}).

\subsubsection{Sources with no LSXPS Counterpart}
\label{sec: nonlsxps}

In addition to those sources described above, we detect $\sim$200 sources in the DGPS mosaics which do not have LSXPS counterparts within $60\arcsec$ \citep{LSXPS}. We refer to these as non-LSXPS sources throughout the manuscript. 
There are a number of plausible reasons as to why these sources would not have been detected in the LSXPS mosaics, including a different combination of observations used to build the mosaics in LSXPS, hot pixels, which are harder to detect in stacked observations, or a lower background to variable or transient sources in the DGPS mosaics as they include less overall observations. Therefore, there is no obvious reason to exclude these sources from our catalog.

After removing sources with field flags or those lying in the PSF of a bright source, we are left with 126 sources, 83 classified as \textit{good} and 43 as \textit{reasonable}. Based on simulations of \textit{Swift}/XRT point sources \citep{Evans2014,Evans2020}, these sources are detected at the 99\% confidence level. 

We utilized the Python API tool to call the Swift-XRT LSXPS Upper limit server\footnote{\url{https://www.swift.ac.uk/LSXPS/ulserv.php}} \citep{LSXPS}, which allows for the calculation of $3\sigma$ upper limits for any position within the LSXPS footprint. 
We specifically called only the DGPS observations covering the position of each source. Aperture photometry using a circular region with a radius of 12 pixels ($28\arcsec$) was then performed on the images to determine the source and background counts in each energy band. We then applied the Bayesian procedure of \citet{Kraft1991} to determine whether the source is detected at the $3\sigma$ level, and, if detected, the mean number of counts and $1\sigma$ errors. The Upper Limit Server also computes a PSF correction to account for vignetting and the encircled energy fraction of the circular aperture. After multiplying the number of counts by this correction factor and dividing by the exposure time, we obtain a count rate in each energy band. This is all done through the \texttt{mergeUpperLimits} tool. These methods are identical to those utilized to compute count rates for LSXPS sources. %\textcolor{blue}{Following this, we compute the hardness ratios $HR_1$ and $HR_2$ following \citet{Park2006}. (do we?}

However, we only find a $3\sigma$ detection for 35 out of 126 sources with 22 detections in the FB, 17 in the HB, 9 in the MB, and 6 in the SB. Of the 35 sources, 16 were detected in multiple bands using this method. This serves to confirm that at least some of these sources, likely more than 35, are not spurious in nature. We note that the \citet{Evans2020} source detection algorithm does not necessarily require a $3\sigma$ statistical significance for detection, and, in fact, the signal-to-noise ratio for many of these sources is $\sim$\,$2$. Instead, the algorithm computes a likelihood that the source is real, which was calibrated using simulations \citep{Evans2014,Evans2020}. This could explain why only 35 of 126 sources are above the $3\sigma$ threshold according to  \citet{Kraft1991}. 

We convert the count rate to an unabsorbed flux ($0.3$\,$-$\,$10$ keV) for each source assuming the median ECF for all DGPS sources detected in LSXPS (\S \ref{sec: finalcat}). This is dependent on the energy band, and we find median values of ECF$_\textrm{FB}=2.7\times 10^{-11}$ erg cm$^{-2}$ cts$^{-1}$ for the full band, and ECF$_{SB}=5.2\times10^{-11}$ erg cm$^{-2}$ cts$^{-1}$, ECF$_{MB}=6.6\times 10^{-11}$ erg cm$^{-2}$ cts$^{-1}$, and ECF$_{HB}=4.5\times 10^{-11}$ erg cm$^{-2}$ cts$^{-1}$. These ECFs were all determined assuming a power-law X-ray spectrum with photon index $\Gamma$\,$=$\,$1.7$ and Galactic hydrogen column density \citep{Willingale2013}.

For these non-LSXPS sources we record only the standard position derived by the source detection algorithm as performed on the DGPS mosaics. We note that these sources have no multi-epoch (i.e., variability) information, as they are only detected in stacked observations (mosaics). Furthermore, due to their faintness and low number of photons, the hardness ratio information is limited, and instead we record clearly the bandpass in which the source is detected. Due to these limitations, we record the non-LSXPS sources in a separate table from those with additional LSXPS information. 
We report the results for these 126 sources in Table \ref{tab: not_in_LSXPS}\footnote{Table \ref{tab: not_in_LSXPS} is available in electronic form at the
CDS via anonymous ftp to cdsarc.u-strasbg.fr (130.79.125.5) or via \url{http://cdsweb.u-strasbg.fr/cgi-bin/qcat?J/ApJS/}.}. We emphasize that these sources, in addition to those in Table \ref{tab: main_cat}, comprise the full DGPS Phase-I catalog.

 \begin{figure} 
\centering
\includegraphics[width=\columnwidth]{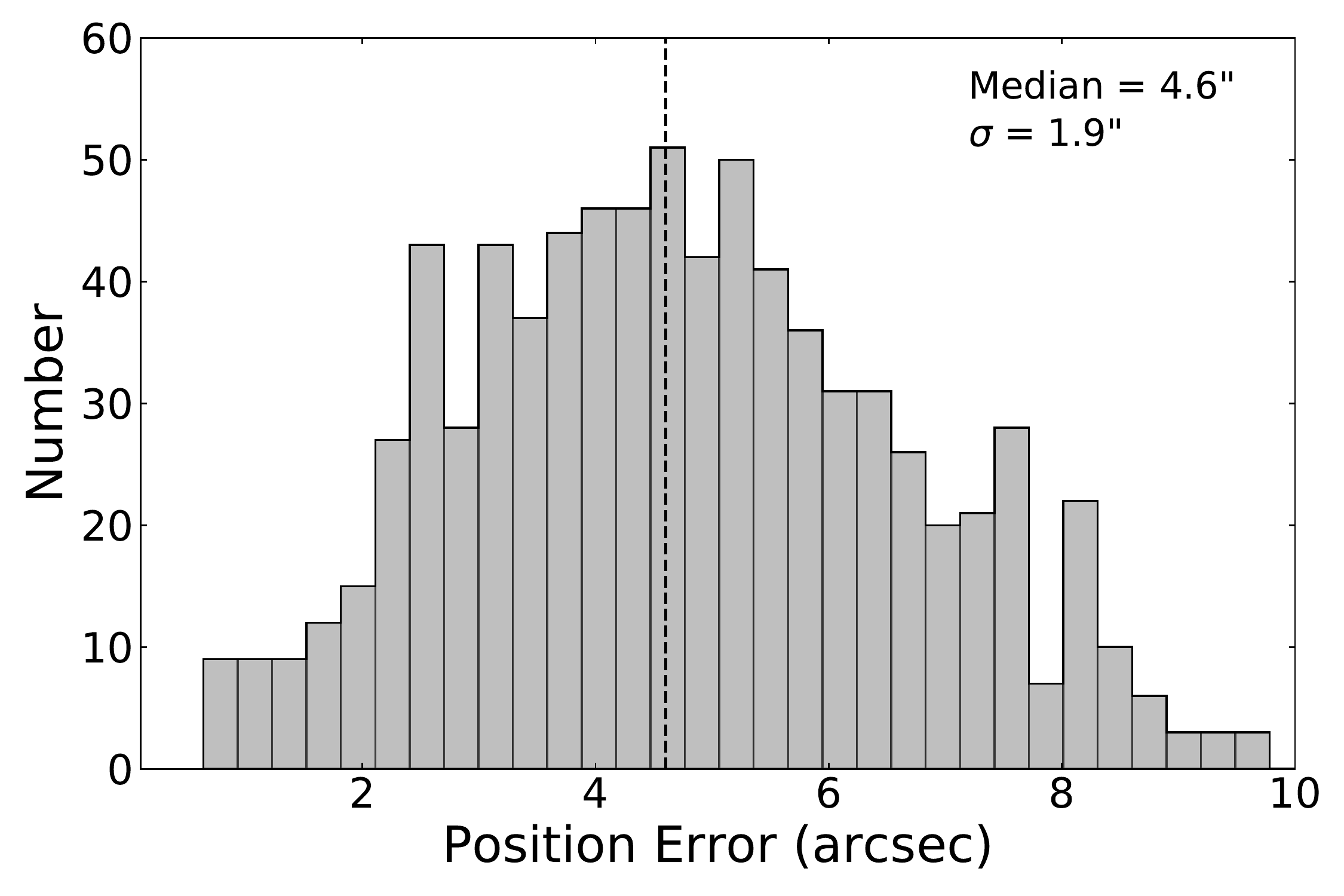}
\includegraphics[width=\columnwidth]{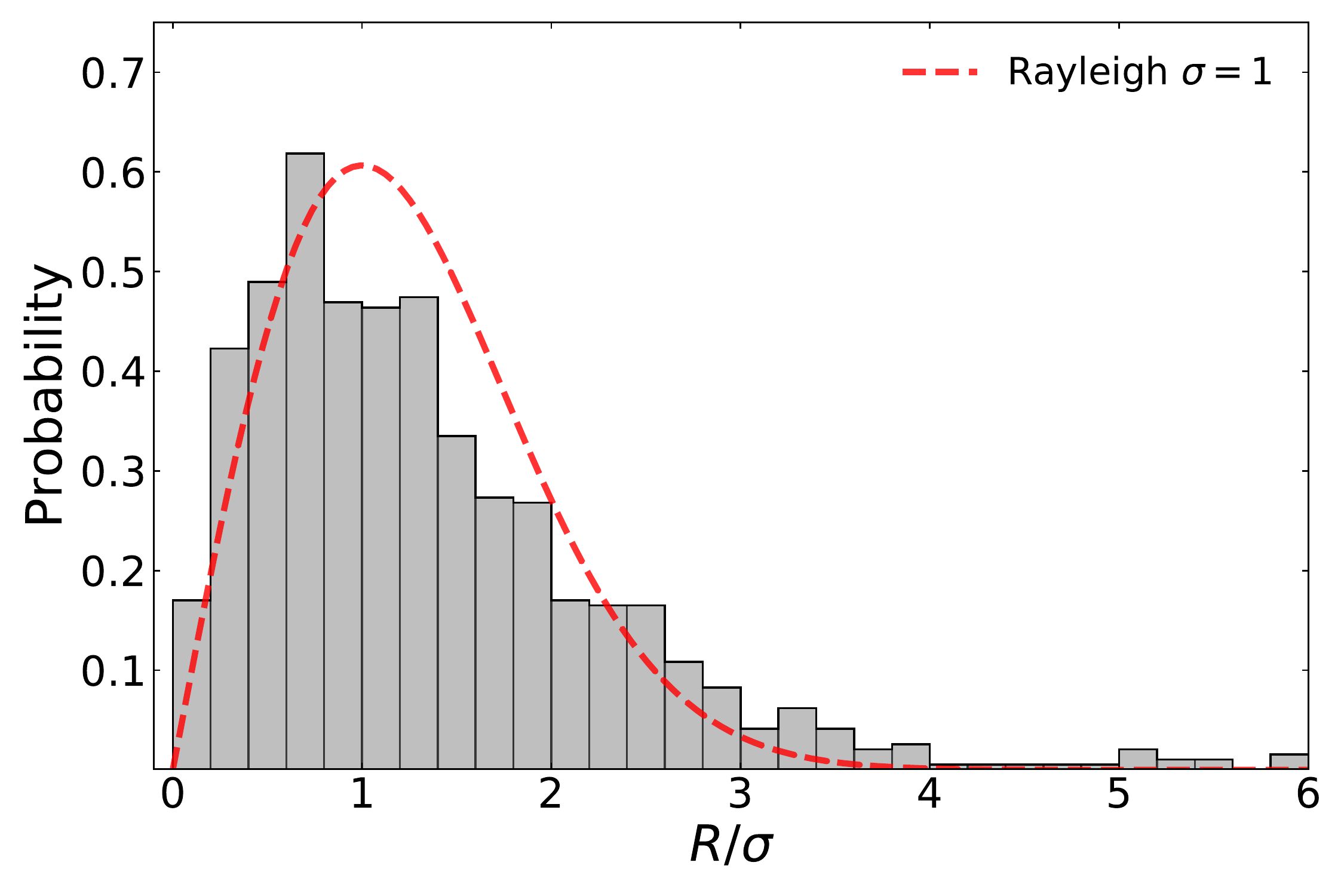}
\caption{
\textbf{Top}: Histogram of the 90\% X-ray  position error for sources in the DGPS catalog. 
\textbf{Bottom}: Radial separation $R$ divided by the 68\% error of the DGPS sources and the 68\% error of other X-ray source error added in quadrature. The radial separation is from the DGPS source to the X-ray counterpart centroid from 4XMM, 2CSC, and 1SWXRT. The dashed red line shows the expected Rayleigh distribution with $\sigma$\,$=$\,$1$. 
}
\label{fig: pos_error_hist}
\end{figure}

\section{Results} 
\label{sec: results}
\subsection{Cross-matching with External Catalogs}
\label{sec: crossmatch}

%\brendan{Very good idea from oleg: to assist in optical/nIR counterpart identification could first cross-match our sources with XMM/Chandra and then take the XMM/Chandra localizations for cross-match with radio/optical/nIR etc to lower spurious fraction and get nice matches!}

We cross-matched the 802 sources in Table \ref{tab: main_cat} with a variety of radio, optical, infrared, and X-ray catalogs in order to identify their multi-wavelength counterparts. We defined a match as when the catalog and DGPS positions were consistent at the 99.7\% confidence level\footnote{In order to convert between the 90\% and 99.7\% position error, we have assumed that our source position errors follow Rayleigh statistics \citep{Evans2014,Evans2020}.} when adding both catalog and DGPS errors in quadrature. The distribution of 90\% position errors are shown in Figure \ref{fig: pos_error_hist} (top panel). The median 90\% position error is $4.6\arcsec$, leading to a 99.7\% error of $\sim$\,$7\arcsec$. 

We began by searching the SIMBAD Astronomical Database \citep{Wenger2000} in order to identify any previous source classifications. As the SIMBAD database does not include positional errors uniformly it is possible some real associations were missed. For all other catalogs, we include the catalog's positional error added to the DGPS position error in quadrature. 

We used \texttt{astroquery} \citep{Ginsburg2019} to search the VizieR Database \citep{Ochsenbein2000} for the following X-ray catalogs: the \textit{Chandra} Source Catalog \citep[CSC;][]{Evans2010} Release 2.0, the \textit{XMM-Newton} Serendipitous Source Catalog \citep[4XMM-DR9;][]{Webb2020,Traulsen2020}, 1SXPS \citep{Evans2014}, and 2SXPS \citep{Evans2020}, 1SWXRT \citep{DElia2013}. In addition to the number of matches in each X-ray catalog we report the number of unique, previously unknown, X-ray sources.  %the \textit{Swift} Galactic Plane Survey \citep{Reynolds2013}, 
We additionally searched the following optical, infrared, and radio catalogs: USNO-B1 \citep{Monet2003}, \textit{Gaia} EDR3 \citep{Gaia2021}, the Two Micron All Sky Survey \citep[2MASS;][]{Skrutskie2006}, and the Very Large Array Sky Survey \citep[VLASS;][]{Lacy2020}. %, and the MeerKAT Galactic Plane Survey \citep[MeerGAL;][]{Thompson2016} Point Source Catalog. 
The results of our cross-matching analysis are displayed in Table \ref{tab: cross-match}. % and Figure \ref{fig: class_pie}. 

We find that 249 (31\%) of DGPS sources were previously unknown to other X-ray surveys (with the exception of LSXPS). In Table \ref{tab: main_cat} we record whether a source has a known X-ray counterpart. Figure \ref{fig: pos_error_hist} (bottom panel) shows the distribution of offsets between X-ray source matches normalized by the 68\% position uncertainty of both sources added in quadrature. The distribution of position-error-normalized offsets approximately follows a Rayleigh distribution with scale parameter $\sigma$\,$=$\,$1$. 
However, there is some excess at $R/\sigma$\,$>$\,$3$ that may hint at an additional systematic position error that was not included. We note that counterparts in 2SXPS are not included in this calculation as their separations are tighter than a Rayleigh distribution due to the use of a similar source detection algorithm on similar data, i.e., the first half of the DGPS data obtained between 2017 and 2019 are included in the 2SXPS catalog. This leads to a bias towards the same centroid location for counterparts in 2SXPS, whereas there is no overlap with 1SWXRT, and, therefore, no bias against a Rayleigh distribution.

We determined the number of false associations by shifting all DGPS sources randomly by $1-2\arcmin$ and repeating the cross-match. All matches found after shifting are considered false positives. 
We repeated this procedure multiple times. 
Due to the high density of optical and infrared sources in the crowded GP, generally there are multiple counterparts within a typical X-ray localization (e.g., between $2-4$ \textit{Gaia} counterparts are found on average for DGPS sources). 
This is reflected in the high false positive fraction ($>$\,$77\%$). Therefore, the determination of the true counterpart is difficult using XRT positions alone. Through our follow-up campaigns, we found that \textit{Chandra} observations were pivotal to the identification of the true multi-wavelength counterpart (see \S \ref{sec: newly classified}). 

\begin{table}
    \vspace{5mm}
    \centering
    \caption{Results of multi-wavelength cross-matching with external catalogs using the combined $3\sigma$ source localization. The expected fraction of spurious matches was determined by shifting our source catalog by $1$\,$-$\,$2\arcmin$ and re-running our cross-matching algorithm. 
    }
    \label{tab: cross-match}
    \begin{tabular}{lcc}
    \hline
    \hline
   External Catalog &  Matches  & Spurious Matches \\
    \hline
    \hline
   \multicolumn{3}{c}{X-ray Catalogs} \\
    \hline
  2CSC & 186 & 2 (1.0\%)  \\
  4XMM-DR9 & 264 & 3 (1.1\%) \\
  2SXPS & 463  & 3 (0.6\%) \\
  1SWXRT & 63  & 1 (1.6\%) \\
  Unique & 249 & -- \\
%  \citet{Reynolds2013} & &  \\
%  \textit{INTEGRAL} & &  \\
%  ASCA & &  \\
%  2RXS & &  \\
\hline 
  \multicolumn{3}{c}{Radio Catalogs} \\
    \hline 
      VLASS & 17 & 1 (5.9\%)  \\
      %\textcolor{red}{MeerKAT} & 57 & \textcolor{red}{XX} \\ 
      Unique & 745 & -- \\ %733 & -- \\ %733 when using meerkat also
  \hline 
  \multicolumn{3}{c}{Optical/nIR Catalogs} \\
  \hline 
  2MASS & 689  & 618 (90.0\%)  \\
  GAIA & 699 & 635 (90.8\%) \\
  USNO-B1 & 562 & 431 (76.7\%)  \\
  Unique & 58 & -- \\
    %GLIMPSE & &  \\
  %WISE & &  \\
  %UKIDSS & &  \\
  %VVV/VVVX & &  \\
    \hline
    \end{tabular}
\end{table}

\subsubsection{Cross-match of non-LSXPS Sources}
\label{sec: non-lsxps_match}

We performed the same cross-matching analysis outlined in \S \ref{sec: crossmatch} on the 126 non-LSXPS sources (Figure \ref{fig: match_pie} and Table \ref{tab: not_in_LSXPS}). We find 17 matches in the X-ray catalogs searched, implying that these sources largely comprise a faint, previously undiscovered population of X-ray sources. Of these 17 matches, 12 were in 4XMM-DR9, 7 in 2SXPS, and 7 in CSC 2.0. The sources with matches in these catalogs are marked in Table \ref{tab: not_in_LSXPS}

We further note that a cross-match of the non-LSXPS sources with SIMBAD results in only 3 classified source matches, and 123 sources without a SIMBAD counterpart. Therefore, a significantly larger fraction of those sources not in LSXPS are previously unknown and unclassified, likely due to their faintness and lower number of counterparts in other X-ray catalogs. %Furthermore, Figure \ref{fig: Flux_hist} shows that the non-LSXPS sources all lie below the 90\% completeness flux, emphasizing the faintness of this source population. 

While only 17 (13\%) of these sources have a known X-ray counterpart, compared to 69\% in of those also detected by LSXPS, this further implies (see also \S \ref{sec: nonlsxps}) that at least some of these non-LSXPS sources are real. 
Moreover, the 7 sources detected in 2SXPS \citep{Evans2020}, but not in the re-analysis for LSXPS \citep{LSXPS}, emphasizes that the combination of specific observations used to create the mosaic is an important factor in the source detection process. %in particular the effect of pixel randomization during the mosaic creation process. 
%This is exactly the effect at play that led to these non-LSXPS sources.

\subsection{Source Classification Breakdown}
\label{sec: classification}

Our cross-match with the SIMBAD database \citep{Wenger2000} resulted in a total of 251 (27\%) previously classified sources.   
However, we found that in some cases the classification was incorrect or incomplete. Thus, while the SIMBAD database provides a useful check as to whether a source is already known (and cross-listings between the same source in other catalogs), it does not provide a robust measure of the number of confidently classified sources in our catalog.

Therefore, we performed an additional search of other external catalogs containing classified source types, including the McGill Online Magnetar Catalog\footnote{\url{https://www.physics.mcgill.ca/~pulsar/magnetar/main.html}} \citep{Olausen2014}, HMXBCAT\footnote{\url{https://heasarc.gsfc.nasa.gov/W3Browse/all/hmxbcat.html}} \citep{Liu2006}, LMXBCAT\footnote{\url{https://heasarc.gsfc.nasa.gov/W3Browse/all/lmxbcat.html}} \citep{Liu2007}, Australia Telescope National Facility (ATNF) Pulsar Catalogue \citep{Manchester2005}, The Million Quasars (Milliquas) v7.2 Catalogue \citep{QSOs},  Symbiotic stars catalogue\footnote{\url{https://heasarc.gsfc.nasa.gov/W3Browse/all/symbiotics.html}} \citep{symbiotic}, the New Online Database of Symbiotic Variables\footnote{\url{http://astronomy.science.upjs.sk/symbiotics/index.html}} \citep{Jaroslav2019}, X-ray catalog of Galactic O stars \citep{Ostars}, Catalog of X-Ray Detected Be Stars\footnote{\url{https://home.gwu.edu/~kargaltsev/XDBS/}} (XDBS; \citealt{CadenBe}), a catalogue of chromospherically active binary stars \citep{Eker2008}, the Open Cataclysmic Variable Catalog\footnote{\url{https://depts.washington.edu/catvar/index.html}} \citep{OpenCV,Jackim2020}, and IP CVs from Koji Mukai's catalog\footnote{\url{https://asd.gsfc.nasa.gov/Koji.Mukai/iphome/catalog/alpha.html} (cross-matched as of 2022 August 31)}. 
%No matches in Caden's Be Star Catalog...
This ensures we probe the majority of known sources within these classes.

\begin{figure} 
\centering
\includegraphics[width=\columnwidth]{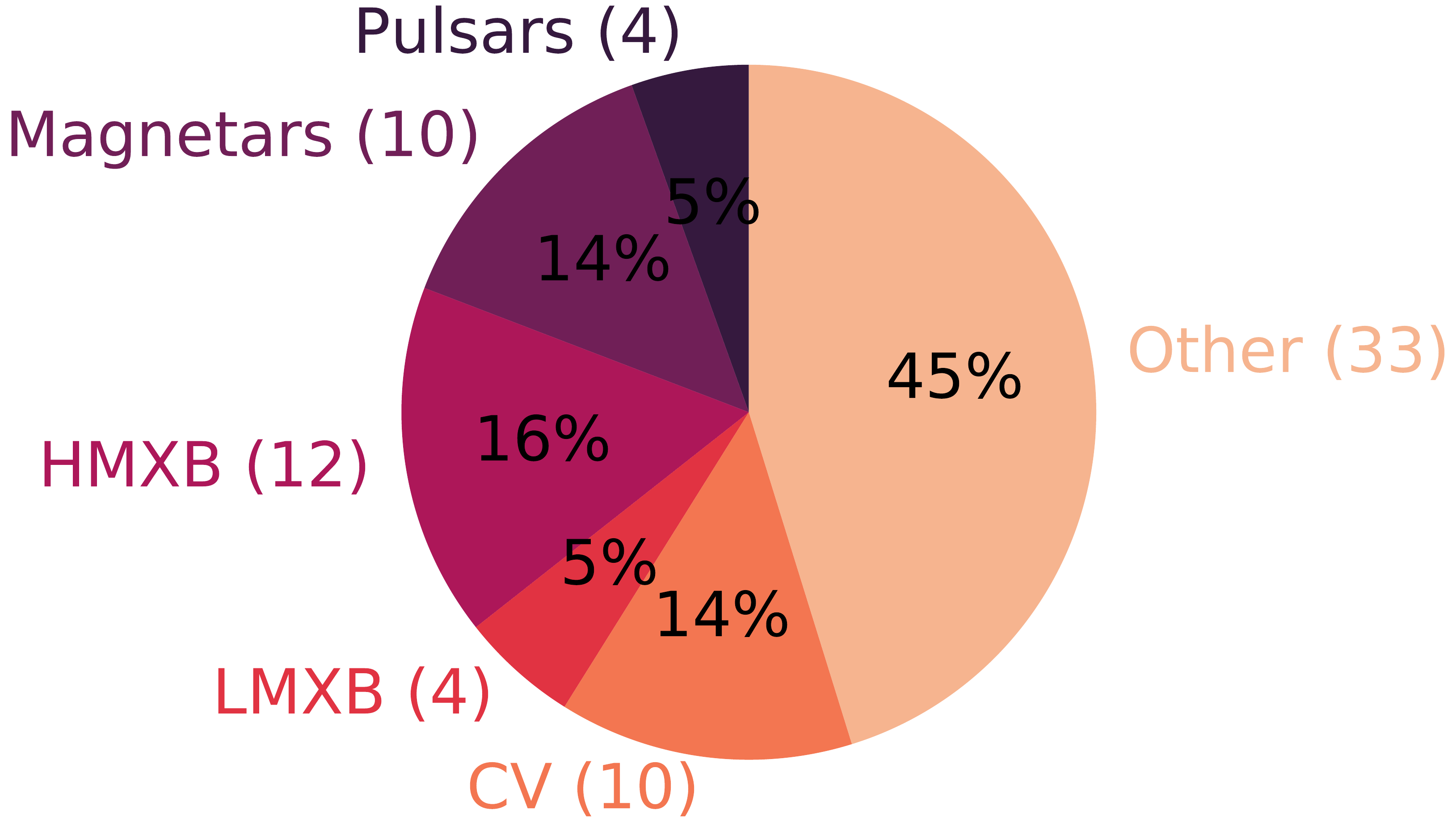}
\caption{Breakdown of the source type for the 73 classified sources in the full DGPS catalog.
}
\label{fig: class_pie}
\end{figure}

In total we find 73 classified sources across the following categories: 
\begin{enumerate}[label=\textit{\roman*})]
    \item 4 pulsars, 
    \item 10 magnetars, 
%14 pulsars (10 being magnetars),
    \item 12 HMXBs, 
    \item 4 LMXBs, 
    \item 10 CVs (6 being IPs), 
    \item 5 Wolf Rayet (WR) stars, 
    \item 18 young stellar objects (YSO),
    \item 5 quasars (QSO), 
    \item 3 symbiotic stars,
    \item and 2 X-ray detected O stars.
\end{enumerate} 
The classification breakdown is demonstrated in Figure \ref{fig: class_pie}. We do not find any associations with X-ray detected Be stars \citep{CadenBe}, %Symbiotic stars \citep{symbiotic,Jaroslav2019}, 
or chromospherically activate binaries \citep{Eker2008}.

Thus, we find only $\sim$\,9\% of DGPS sources are confidently classified. This is likely a lower limit to the true number of classified sources in the Survey given that many of the catalogs searched are over a decade old and may be lacking in completeness. This further emphasizes the need for up-to-date catalogs of source classifications and for machine learning techniques to determine preliminary source classifications for large datasets \citep{Yang2021,Yang2022,Tranin2022}, see \S \ref{sec: machinelearning}. 

In Figure \ref{fig: Flux_hist} we display the X-ray flux distribution of DGPS sources compared to known IP CVs, HMXBs, LMXBs, and magnetars. The large majority of DGPS sources lie below the distribution of classified sources, emphasizing the difficulty in classifying faint sources. This may suggest that the DGPS population of sources could lie at further distances (leading to a lower observed flux), and are, therefore, possibly more absorbed, due to a larger Galactic column density.

\begin{figure} 
\centering
\includegraphics[width=\columnwidth]{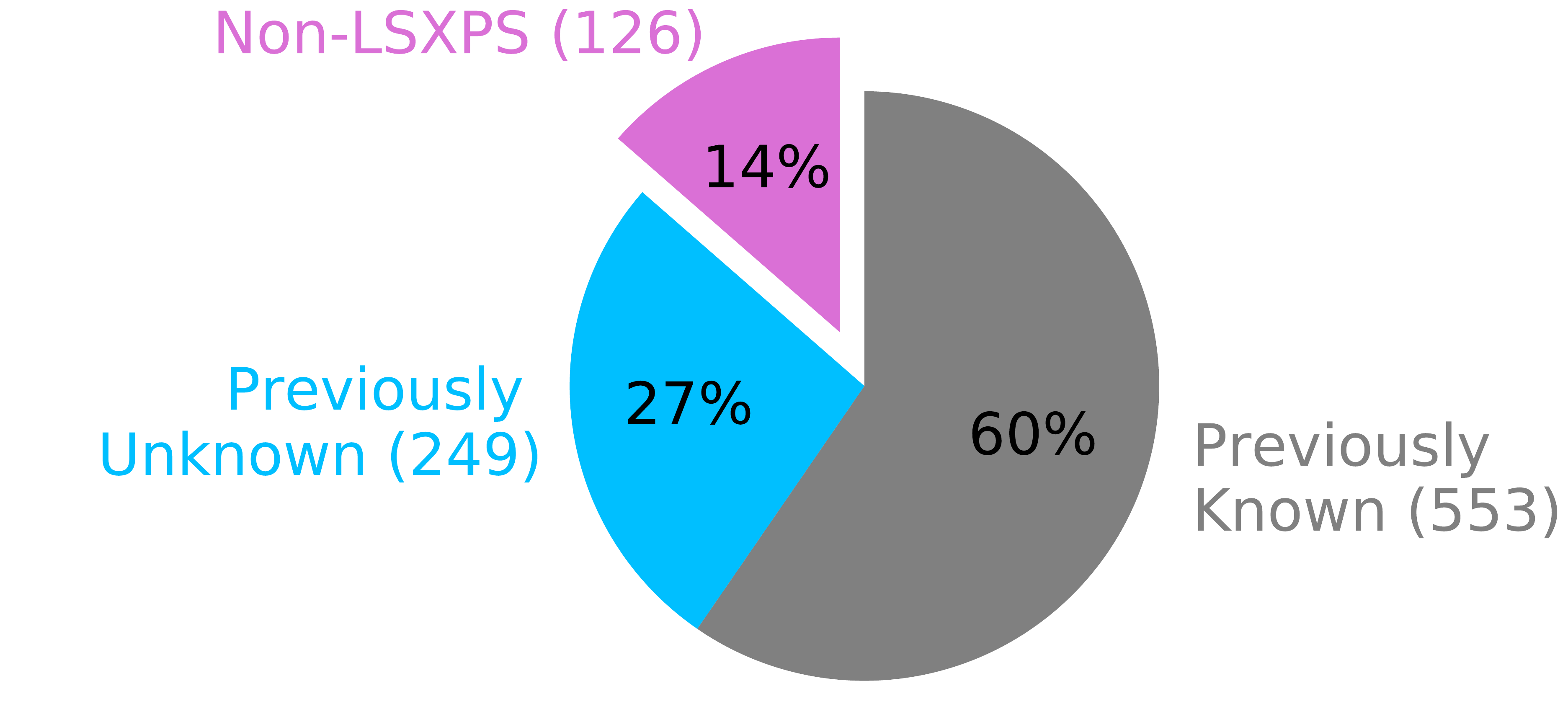}
\caption{Breakdown of the 928 X-ray sources (802 LSXPS $+$ 126 non-LSXPS) in the full DGPS source catalog.}
\label{fig: match_pie}
\end{figure}

\begin{figure} 
\centering
\includegraphics[width=\columnwidth]{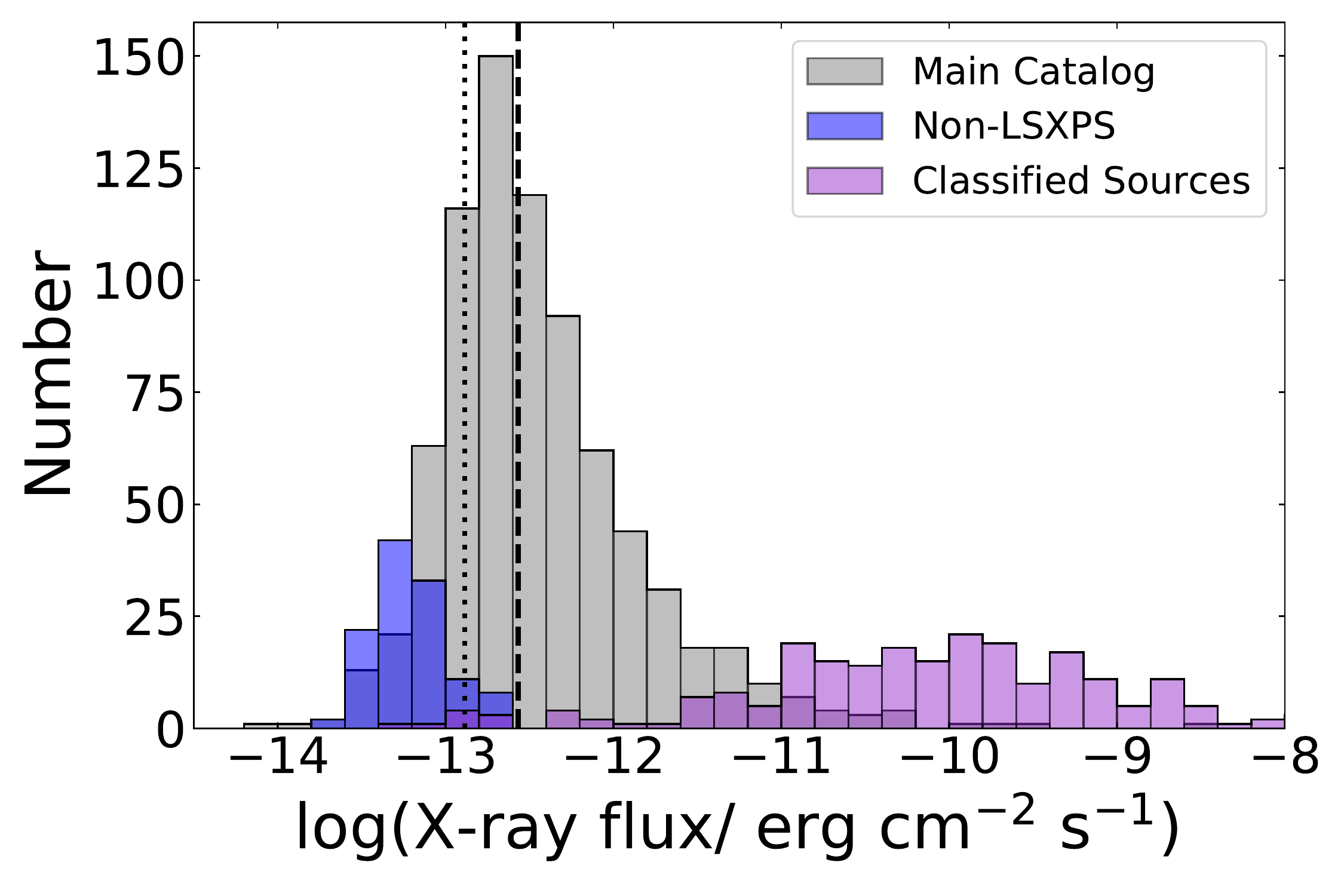}
\caption{Histogram of average flux values for DGPS sources (gray), the non-LSXPS sources (blue), and known classified sources (purple), including IP CVs, LMXBs, HMXBs, and Magnetars. 
The dotted and dashed lines correspond to the 50\% and 90\% completeness flux of the DGPS, respectively (see \S \ref{sec: completeness}).
} 
\label{fig: Flux_hist}
\end{figure}

\begin{figure*} 
\centering
\includegraphics[width=1.8\columnwidth]{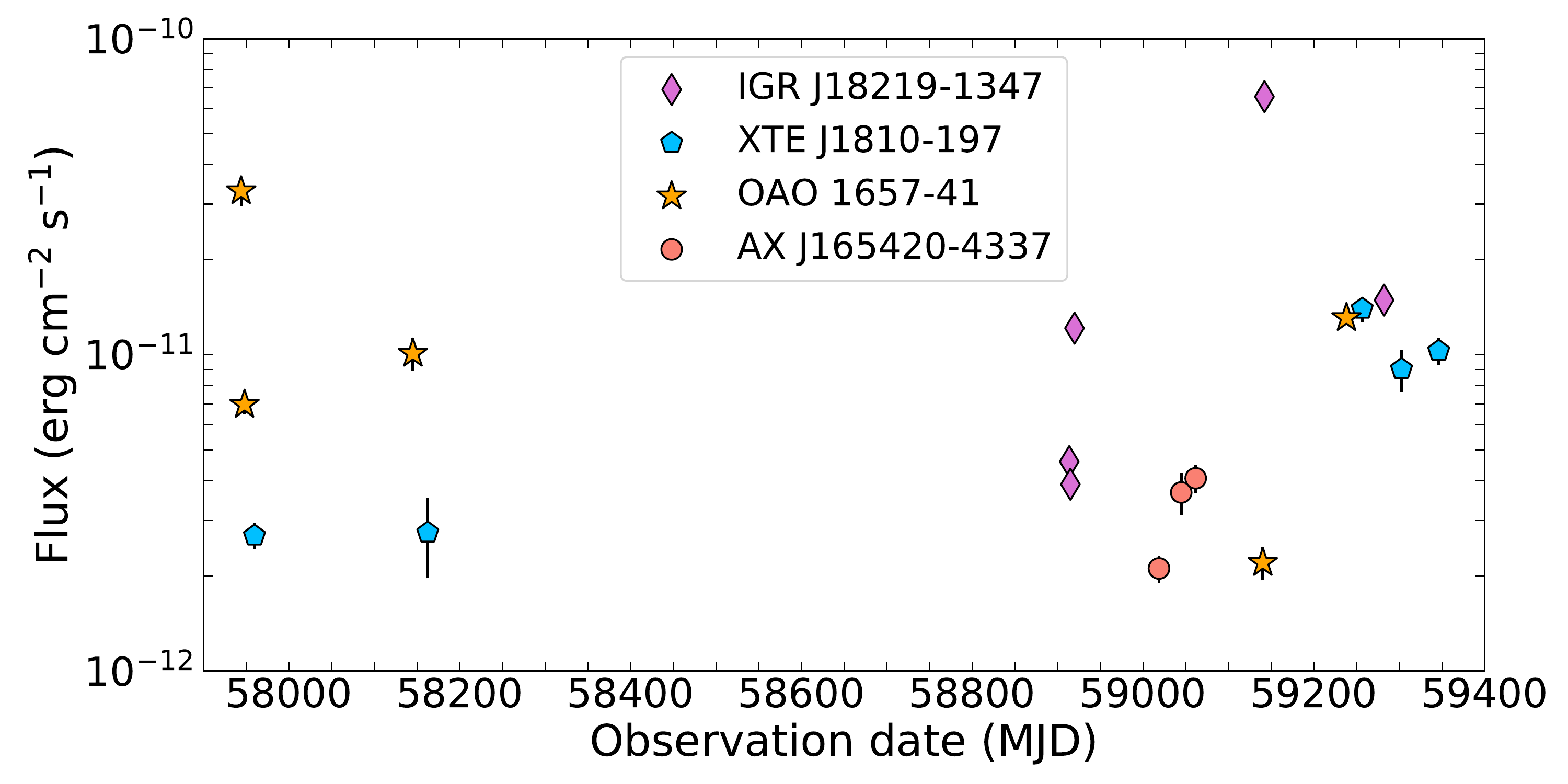}
\caption{Examples of variable sources identified in DGPS observations: IGR J18219-1347 is a BeXRB \citep{OConnor2021}, XTE J1810-197 is a magnetar candidate \citep{Markwardt2003,Israel2004}, OAO 1657-41 is a HMXB \citep{OAO1978,OAO1993}, and AX J165420-43337 (also known as 1RXS J165424.6-433758) is a  polar CV \citep{OConnor2023polar}. 
} 
\label{fig: lightcurve_ex}
\end{figure*}

\subsection{Variable X-ray Sources}
\label{sec: variable}

The DGPS was aimed at uncovering new or variable X-ray sources within the GP. This was done through the rapid analysis of quick-look data (\S \ref{sec: quicklook}) and the comparison of source flux levels with archival observations. An example of variable sources uncovered in DGPS observations is displayed in Figure \ref{fig: lightcurve_ex}. The majority of sources displaying obvious variable behavior were already classified (typically HMXBs, LMXBs, or magnetars; Figure \ref{fig: lightcurve_ex}), but we were also able to classify a number of variable sources (e.g., \citealt{Gorgone2019,Gorgone2021,OConnor2021,OConnor2023polar,OConnor2023IP}) through our follow-up programs, with more classifications in progress.

\begin{figure} 
\centering
\includegraphics[width=\columnwidth]{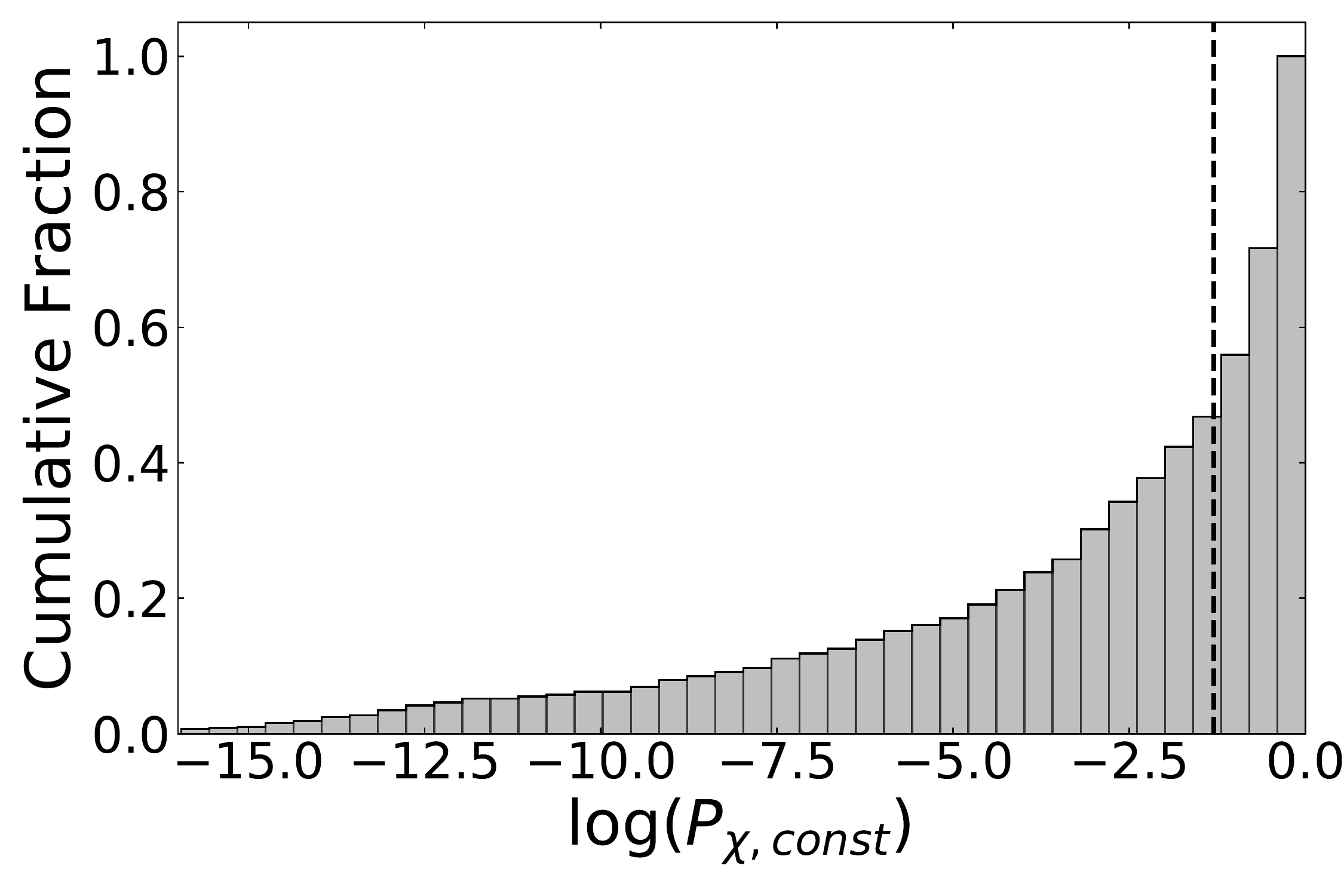}
\caption{Cumulative distribution of the Pearson's $\chi^2$ variability test for all DGPS sources. The dashed line represents a threshold of $P_{\chi,\textrm{const}}=0.05$, below which a source is considered variable. Approximately $50\%$ of sources lie below this threshold.
}
\label{fig: pearson}
\end{figure}

For the purposes of the DGPS catalog, we make use of the Pearson's $\chi^2$ variability test (see also \citealt{Evans2014,Evans2020}). This test computes the probability that the source count rate is constant across all \textit{Swift} observations of the source. %The smallest numerical value returned by this test is $10^{-16}$, which is consistent with a probability of zero that the source is constant.
We consider a source variable if the probability is $P_{\chi,\textrm{const}}$\,$<$\,$0.05$. Approximately half of DGPS sources are expected to display variability (i.e., they are not constant) with a probability higher than 95\% (Figure \ref{fig: pearson}).

\begin{figure} 
\centering
\includegraphics[width=\columnwidth]{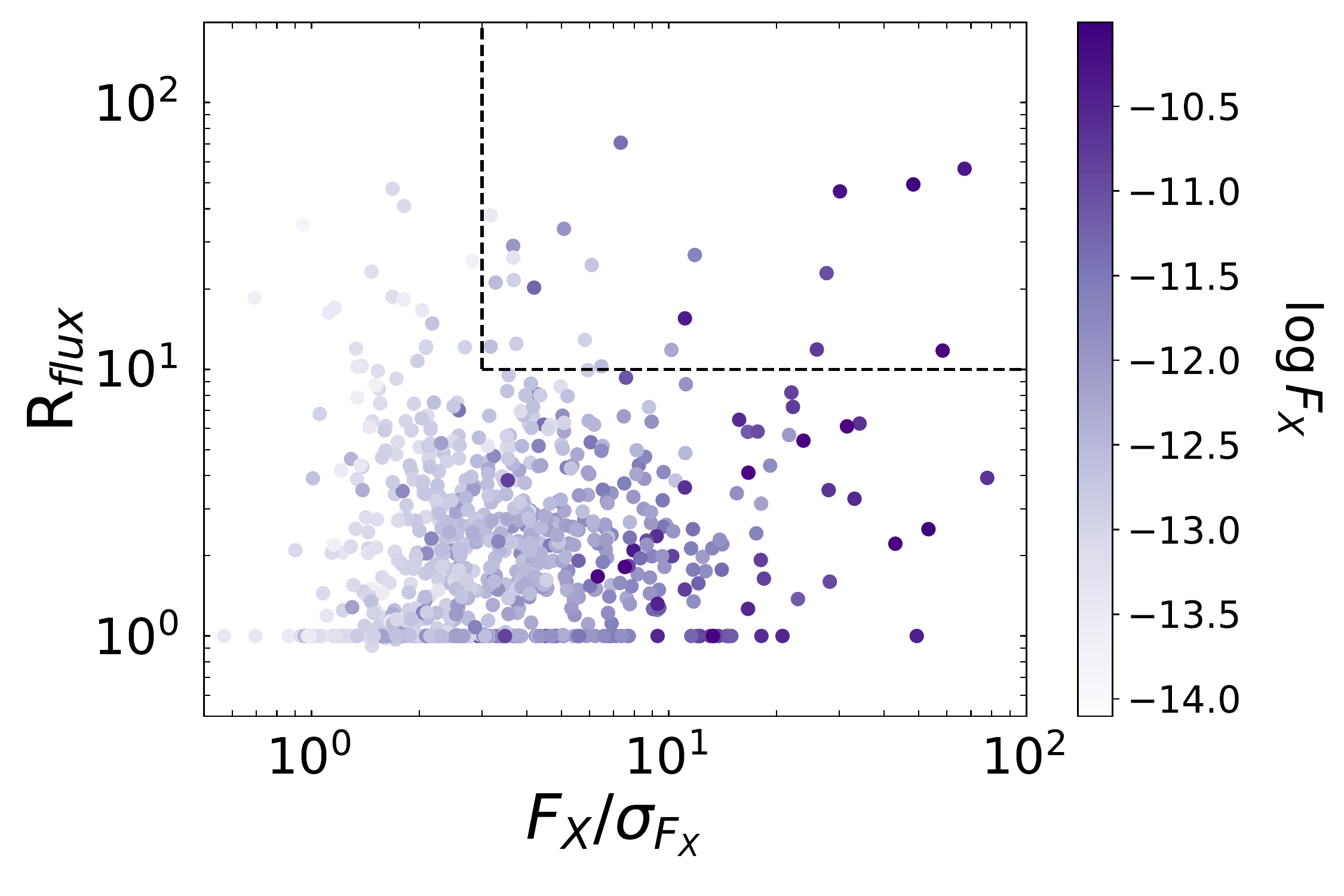}
\caption{The ratio of the peak-to-mean X-ray flux $R_\textrm{flux}$ versus the ratio of the mean X-ray flux $F_X$ and the X-ray flux error $\sigma_{F_X}$. The points are colored by the log of the $0.3$\,$-$\,$10$ keV X-ray flux in erg cm$^{-2}$ s$^{-1}$. The black dashed line indicates a region of parameter space where sources are likely flaring or highly variable.
}
\label{fig: variability_vs_brightness}
\end{figure}

In addition, following \citet{Eyles2022}, we compute the ratio of the peak-to-mean X-ray flux, denoted by $R_\textrm{flux}$, as an indicator of flaring sources. We display $R_\textrm{flux}$ for each source in Figure \ref{fig: variability_vs_brightness}. We find that only 50 sources in the Survey are consistent with $R_\textrm{flux}$\,$>$\,$10$ and 138 with $R_\textrm{flux}$\,$>$\,$5$. 
%22 out of 50 sources have a known x-ray match and 6 are classified
Out of the 50 sources with $R_\textrm{flux}$\,$>$\,$10$, only 31 satisfy $F_X/\sigma_{F_X}$\,$>$\,$3$ (Figure \ref{fig: variability_vs_brightness}). Thus only 31 of these sources have accurate enough flux determinations that the increase in flux by an order of magnitude is statistically significant.

If we further sort these to sources with $F_X$\,$>$\,$10^{-12}$ erg cm$^{-2}$ s$^{-1}$, our threshold for source follow-up (\S \ref{sec: quicklook}), we find that only 11 sources satisfy these criterion, all of which are classified and have a known X-ray counterpart: 1 LMXB, 4 HMXBs, 3 magnetars, 1 pulsar, a pulsar wind nebula \citep{Ng2008},  %https://ui.adsabs.harvard.edu/abs/2008ApJ...686..508N/abstract
and the young star cluster Westerlund 1. 
This is in contrast to a total of 151 sources with $F_X$\,$>$\,$10^{-12}$ erg cm$^{-2}$ s$^{-1}$ in the DGPS catalog (115 of which have a known X-ray counterpart). %In any case, the fact that these sources are classified as HMXBs and magnetars (including a LMXB already classified by the DGPS; \citealt{Gorgone2019}), suggests that our criteria for identifying these interesting sources is valid. 

\section{Discussion}
\label{sec: discussion}

\subsection{Completeness}
\label{sec: completeness}

We estimated the completeness of the DGPS catalog using the simulations performed by \citet{Evans2014,Evans2020}. %This method was previously used by \citet{Kennea2018}. 
\citet{Evans2014,Evans2020} performed detailed simulations of source detection likelihood with \textit{Swift}/XRT as a function of flux and exposure time. The source detection algorithm utilized in this work is most similar to \citet{Evans2020}, which displayed a factor of $3.5\times$ improvement in sensitivity compared to \citet{Evans2014} due to differences in the detection procedure and a more accurate modeling of the XRT PSF. Therefore, we estimate our completeness using Figure 6 of \citet{Evans2020}. We used the simulations corresponding to the inclusion of sources classified as both `\textit{good}' and `\textit{reasonable}'.

The median exposure time of DGPS tiles is $\sim$4.6 ks. Using the calculations performed by \citet{Evans2020}, this corresponds to a 50\% completeness flux of $1.3\times10^{-13}$ erg cm$^{-2}$ s$^{-1}$ and a 90\% completeness of $2.7\times10^{-13}$ erg cm$^{-2}$ s$^{-1}$. However, as shown in Figure \ref{fig: expmap_mosaic}, the exposure varies over the GP due to regions of overlap between tiles. Therefore, these completeness values may underestimate the true fraction of faint sources expected in the overlap regions (see Figure \ref{fig: expmap_mosaic}).

In order to account for this, we performed a Monte Carlo simulation to sample exposure times from random locations within the Survey footprint (Figure \ref{fig: expmap_mosaic}). We then estimated the 50\% and 90\% completeness using the same method outlined above. We repeated this procedure for 20,000 locations in order to find a distribution of completeness flux levels across the Survey. We find a 50\% completeness flux of $(1.3^{+0.3}_{-0.4})\times10^{-13}$ erg cm$^{-2}$ s$^{-1}$ and a 90\% completeness of $(2.7^{+0.4}_{-0.7})\times10^{-13}$ erg cm$^{-2}$ s$^{-1}$. As expected, these values are consistent with our initial estimate.

 \begin{figure} 
\centering
\includegraphics[width=\columnwidth]{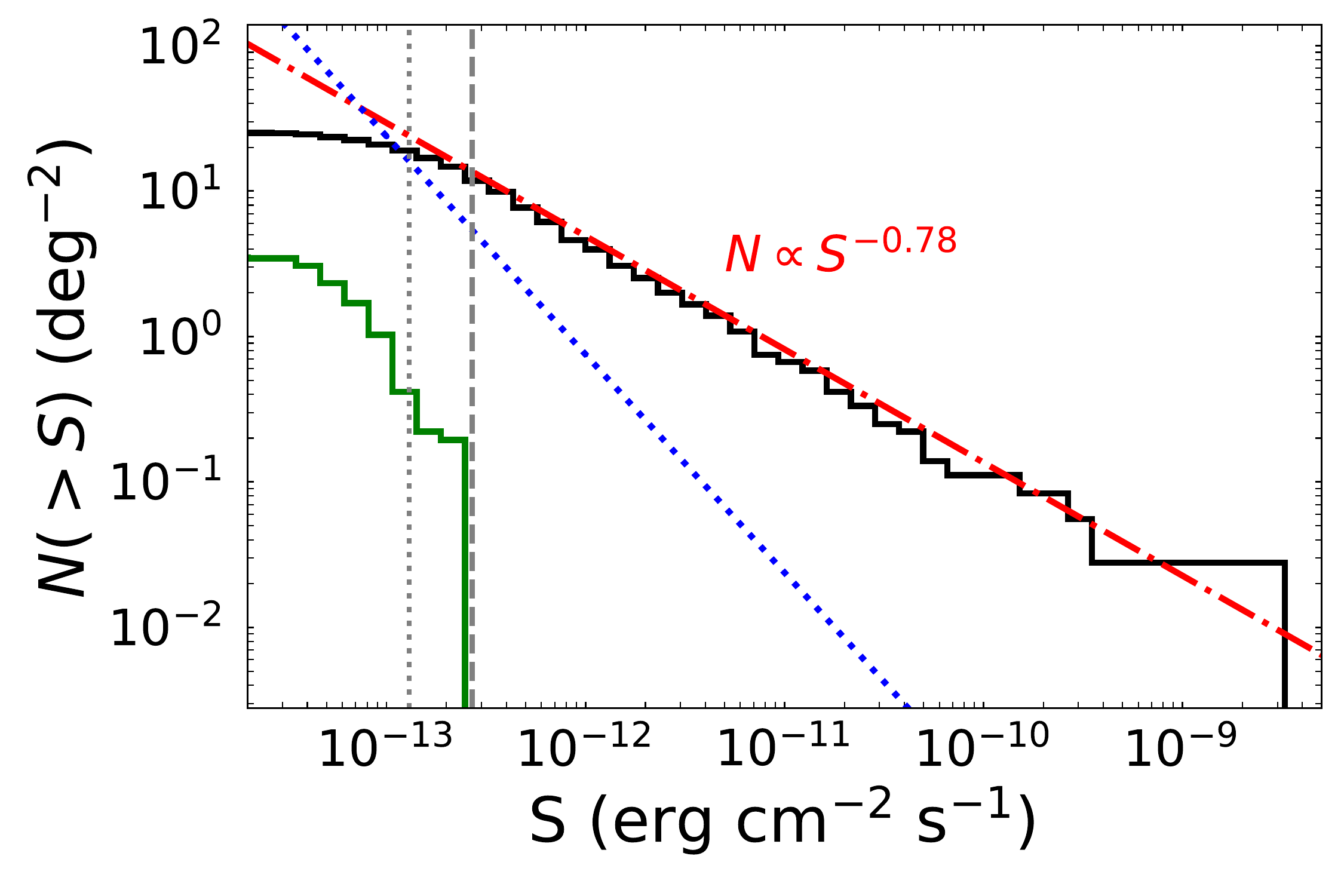}
\caption{$\log N$\,$-$\,$\log S$ plot for the full \textit{Swift} DGPS ($0.3$\,$-$\,$10$ keV) catalog. The best fit line is displayed in red corresponding to $N(>S)\propto S^{-0.78}$. The green solid line displays the $\log N$\,$-$\,$\log S$ for non-LSXPS sources.
The dotted and dashed lines correspond to the 50\% and 90\% completeness flux of the DGPS, respectively. The blue dotted line is an estimate of the extragalactic source population \citep{Ueda1999}. 
}
\label{fig: logN-logS}
\end{figure}

\subsection{Luminosity Function}
\label{sec: lum func}

%\textcolor{blue}{compare to https://iopscience.iop.org/article/10.1086/307291/pdf and  https://articles.adsabs.harvard.edu/pdf/2004MNRAS.351...31H and https://www.aanda.org/articles/aa/pdf/2010/15/aa13570-09.pdf and https://arxiv.org/pdf/astro-ph/0101093.pdf and https://www.aanda.org/articles/aa/pdf/2002/33/aah3162.pdf and https://www.aanda.org/articles/aa/abs/2005/44/aa3513-05/aa3513-05.html}

Using the full DGPS source catalog, we derive the slope and normalization of the $\log N$\,$-$\,$\log S$ curve at Galactic latitudes $|b|$\,$<$\,$0.5$ (Figure \ref{fig: GP_mosaic}). We adopt a power-law form of this curve as $N(>S)$\,$=$\,$K S^\alpha$, where $K$ is a normalization factor. The slope of this
curve yields insight into the spatial distribution of X-ray source populations within our Galaxy. 

In Figure \ref{fig: logN-logS}, we display the $\log N$\,$-$\,$\log S$ derived from the mean fluxes of DGPS sources in the $0.3$\,$-$\,$10$ keV energy range in units of erg cm$^{-2}$ s$^{-1}$. The best fit power-law distribution has a slope $\alpha$\,$=$\,$-0.78\pm0.03$. We have only fitted the distribution for fluxes above the 50\% completeness value (dotted line in Figure \ref{fig: logN-logS}), where the curve rapidly flattens.
%HMXBs $\alpha=-0.6$ \citep{Grimm2002}
We note that including the non-LSXPS sources (\S \ref{sec: nonlsxps}) has no impact on the value of the slope as they all lie below the completeness flux value.

Our value is similar to the slope derived with the ASCA GP Survey \citep{Sugizaki2001} of $-0.79\pm0.07$, and consistent with the $-0.64\pm0.15$ slope derived for HMXBs \citep{Grimm2002}. Both values are flatter than the $-1$ expected for a uniform infinite-plane source distribution. However, past X-ray surveys using different instruments have found values in agreement with $\alpha$\,$\approx$\,$-1$ \citep{Hertz1984,Dean2005}.
These differences may be due to the survey area covered, with different populations of X-ray sources probed, as well as instrument sensitivity. The DGPS covers regions of the plane dominated by spiral arms (Figure \ref{fig: MWfig}) at low Galactic latitudes, and therefore we would expect a shallow slope for the $\log N$\,$-$\,$\log S$ relation  \citep{Sugizaki2001,Grimm2002}, whereas past Galactic X-ray surveys also covered larger scale heights, leading to a steeper slope. We note that the $\log N$\,$-$\,$\log S$ curve for extragalactic X-ray sources is considerably steeper ($\alpha$\,$\approx$\,$-1.5$; \citealt{Gioia1990, Hasinger1993,Ueda1999,Luo2017}), and in agreement with the expectations for a 3D Euclidean Universe ($N$\,$\propto$\,$S^{-3/2}$). %see here for the Euclidean derivation https://ui.adsabs.harvard.edu/abs/1971swng.conf..655H/abstract

In order to determine whether extragalactic sources visible through the plane were contaminating our sample, we estimated their contribution following the methods of \citet{Sugizaki2001} and by converting the $\log N$\,$-$\,$\log S$ fit ($2$\,$-$\,$10$ keV) from \citet{Ueda1999} to the $0.3$\,$-$\,$10$ keV flux, assuming an extragalactic source spectrum with power-law photon index $\Gamma$\,$=$\,$2$ absorbed by $N_H$\,$=$\,$5\times 10^{22}$ cm$^{-2}$. These values were chosen under the assumption that the extragalactic source population comprises only active galactic nuclei (AGN). %We then converted the absorbed $0.3$\,$-$\,$10$ keV flux to the $0.3$\,$-$\,$10$ keV XRT count rate for comparison with our observed distribution. 
The extragalactic population begins to significantly contribute at fluxes less than $10^{-12}$ erg cm$^{-2}$ s$^{-1}$, and has a negligible impact on the population of brighter sources.

\subsection{Catalog Characteristics}
\label{sec: catalog properties}

%for x-ray to optical/infrared ratios see https://arxiv.org/pdf/2110.06234.pdf Eqn 2/3. probably also want to do x-ray to radio ratio for the sources with meerkat counterparts.

\begin{figure*} 
\centering
\includegraphics[width=\columnwidth]{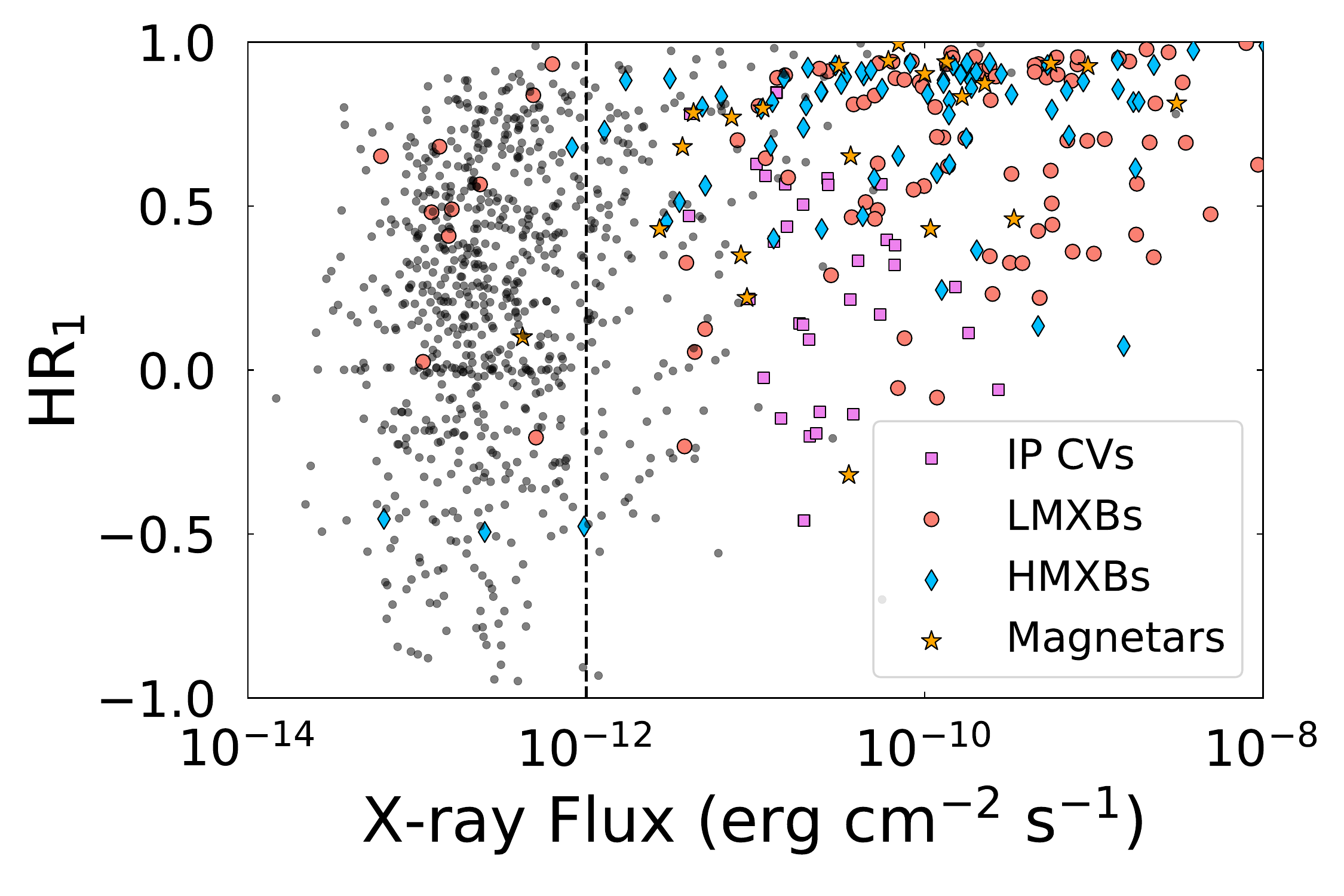}
\includegraphics[width=\columnwidth]{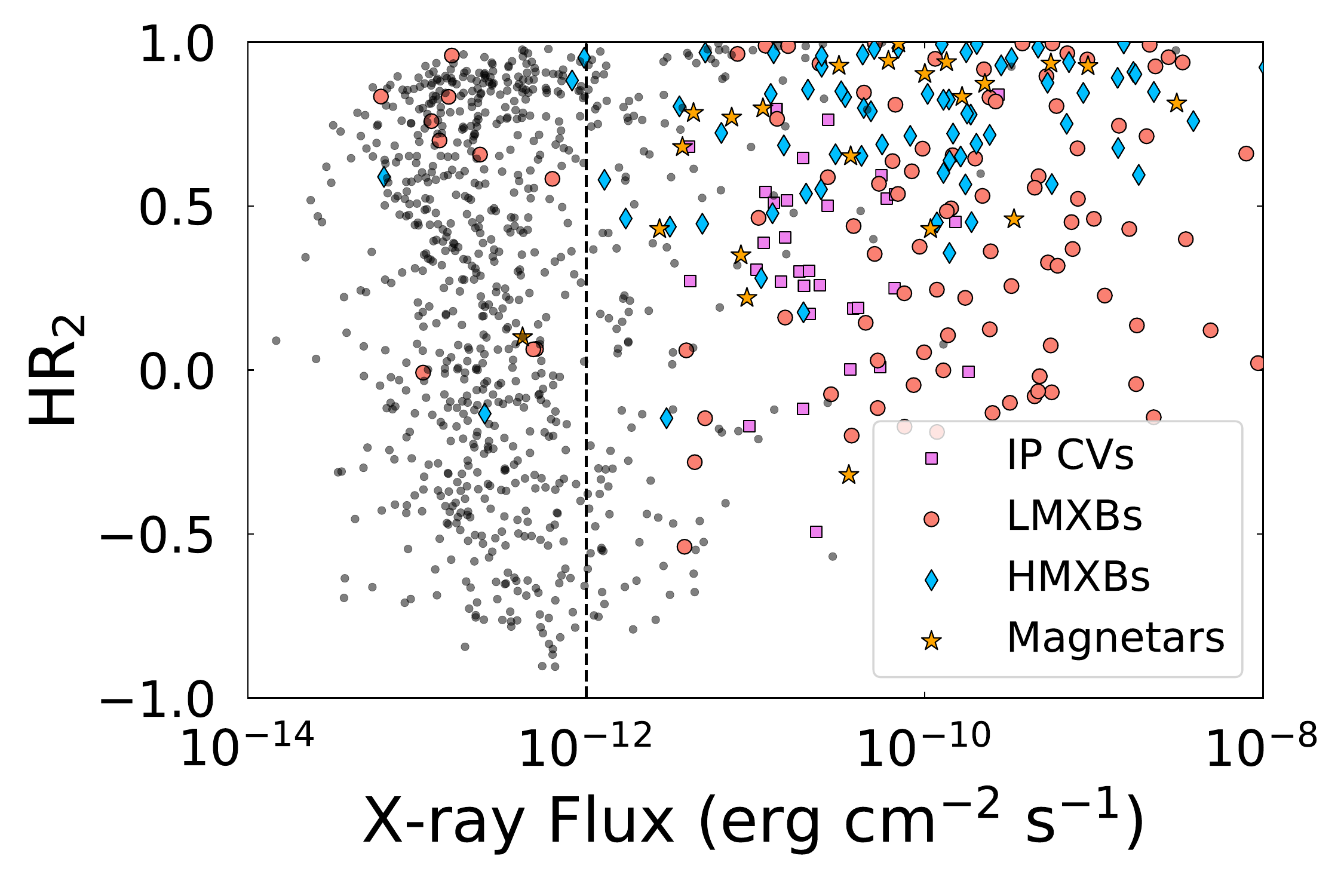}
\caption{Distribution of DGPS sources (gray circles) in terms of hardness ratio and X-ray flux. For reference, we display LMXBs, IP CVs, HMXBs, and magnetars from the 2SXPS catalog.}
\label{fig: HR_vs_Flux}
\end{figure*}

%In Figure \ref{fig: rate_CDF} we display the cumulative distribution of count rates in these energy bands, while Figure \ref{fig: HB_vs_SB_rate} compares the count rate between the HB and SB for sources detected in both bands. The SB and MB rate are typically lower than the HB and FB, which may be due to either absorption by the interstellar medium (ISM) or the smaller energy range covered by those bands. 
Figure \ref{fig: HR_vs_Flux} shows the $0.3$\,$-$\,$10$ keV X-ray flux versus $HR_1$ and $HR_2$ for DGPS sources. For comparison we display the known population of IP CVs, LMXBs \citep{Liu2007}, HMXBs \citep{Liu2006}, and magnetars \citep{Olausen2014} from the 2SXPS catalog (see Appendix \ref{sec: HR derivation} for details).  We see that the majority of our sources lie both below the completeness values (vertical lines) and below the flux of classified sources (Figure \ref{fig: Flux_hist}), underscoring a very large population of faint, unclassified sources. However, it is difficult to classify these sources based on hardness ratios alone, as demonstrated by Figure \ref{fig: HR-plane} (for details see Appendix \ref{sec: HR derivation}). 
There is significant overlap in the population of classified sources, emphasizing the need for machine learning to disentangle source properties in higher dimensional space \citep{Yang2022,Tranin2022}. 

The DGPS sources are distributed relatively uniformly across Galactic longitude (Figures \ref{fig: hist_lat_lon} and \ref{fig: l_vs_b}) within the Survey footprint (\S \ref{sec: footprint}). For example, the number of sources between $10$\,$<$\,$l$\,$<$\,$30$ deg and $330$\,$<$\,$l$\,$<$\,$350$ deg is 413 and 389, respectively. 
However, pockets of longitude with less sources exist. We find that this is due, at least in part, to sources of intense stray light (Figure \ref{fig: GP_mosaic}) at $l$\,$\approx$\,$338$\,$-$\,$342$ deg and $l$\,$\approx$\,$12$\,$-$\,$14$ deg (see the black star in Figure \ref{fig: l_vs_b}; bottom panel). This is caused by the fact that we excluded sources with an LSXPS field flag indicating that they reside in regions of stray light, and, therefore, may be the result of unreliable detections (\S \ref{sec: finalcat}). % This is caused by the fact that we excluded sources with LSXPS field flags notifying us that they reside in regions of stray light as they are, therefore, possibly unreliable detections (\S \ref{sec: finalcat}).
In Galactic latitude we see a marked decrease in sources as we move away from the GP, as expected. In Appendix \ref{sec: appendix_galactic_coord_figs}, we display additional characteristics of sources across the GP (e.g., hardness ratios and variability).

% \begin{figure} 
%\centering
%\includegraphics[width=\columnwidth]{figs/Rate_CDF_all.pdf}
%\includegraphics[width=\columnwidth]{figs/Rate_CDF_all_density.pdf}
%\caption{Cumulative distribution of XRT count rates for sources detected in each of the four energy bands.
%}
%\label{fig: rate_CDF}
%\end{figure}

%\begin{figure} 
%\centering
%\includegraphics[width=\columnwidth]{figs/rate_fig_HB_vs_SB.pdf}
%\caption{Source count rates in the HB ($2$\,$-$\,$10$ keV) versus the SB ($0.3$\,$-$\,$1$ keV) Only sources %with either a blind or retrospective detection detected in both bands are shown. Error bars are not displayed. The dashed line shows sources where the count rate is the same in both bands. }
%\label{fig: HB_vs_SB_rate}
%\end{figure}

\begin{figure*} 
\centering
\includegraphics[width=2\columnwidth]{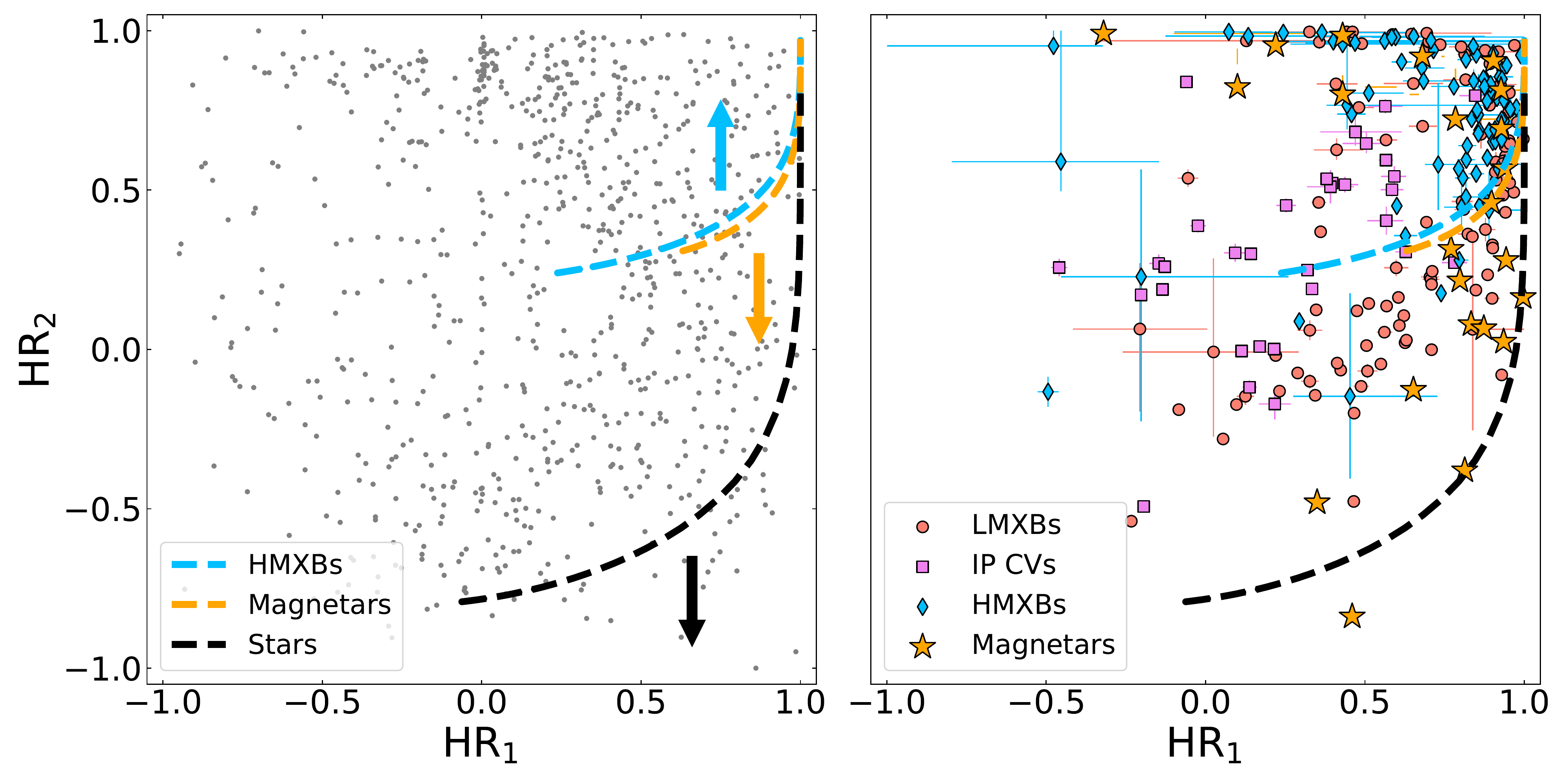}
\caption{\textbf{Left:} Location of DGPS sources (gray circles) in the $HR_1$\,$-$\,$HR_2$ plane. The dashed lines represent typical spectra for HMXBs, Magnetars, and stars as outlined in Appendix \ref{sec: HR derivation}. The arrows mark the general location of these sources with respect to the lines. 
\textbf{Right:} The mean hardness ratios of LMXBs, IP CVs, HMXBs, and magnetars from 2SXPS are shown for comparison. 
}
\label{fig: HR-plane}
\end{figure*}

 \begin{figure} 
\centering
\includegraphics[width=\columnwidth]{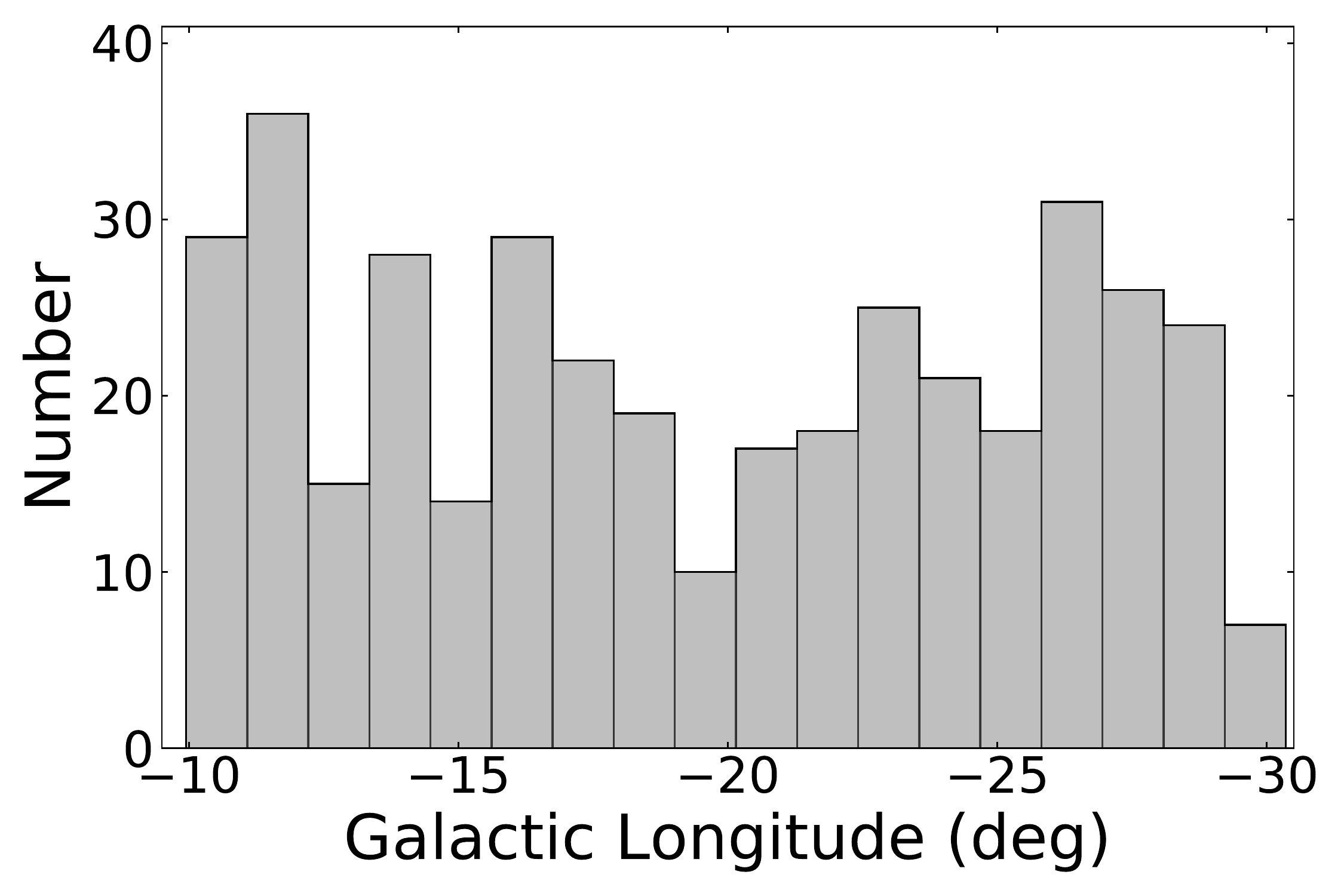}
\includegraphics[width=\columnwidth]{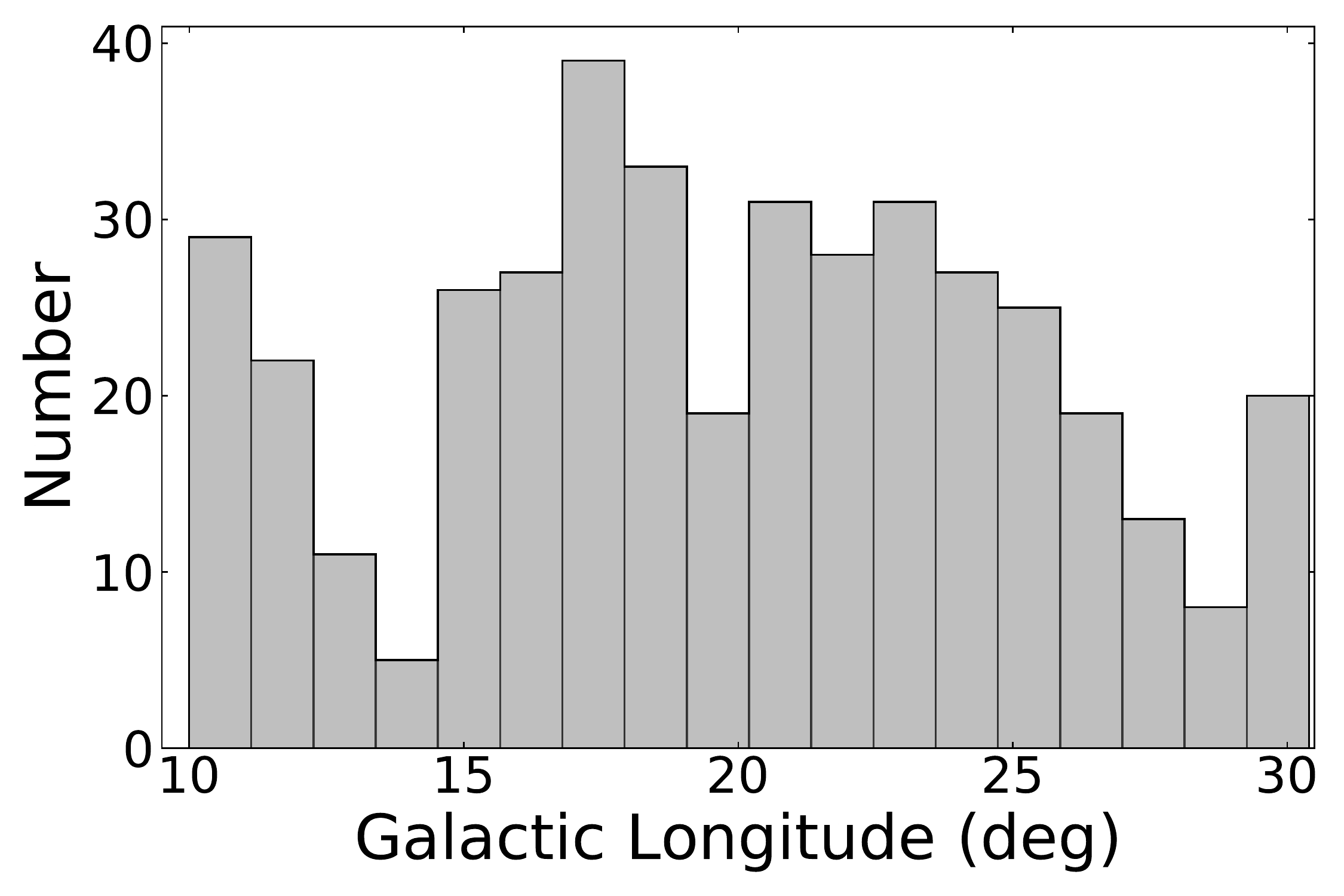}
\includegraphics[width=\columnwidth]{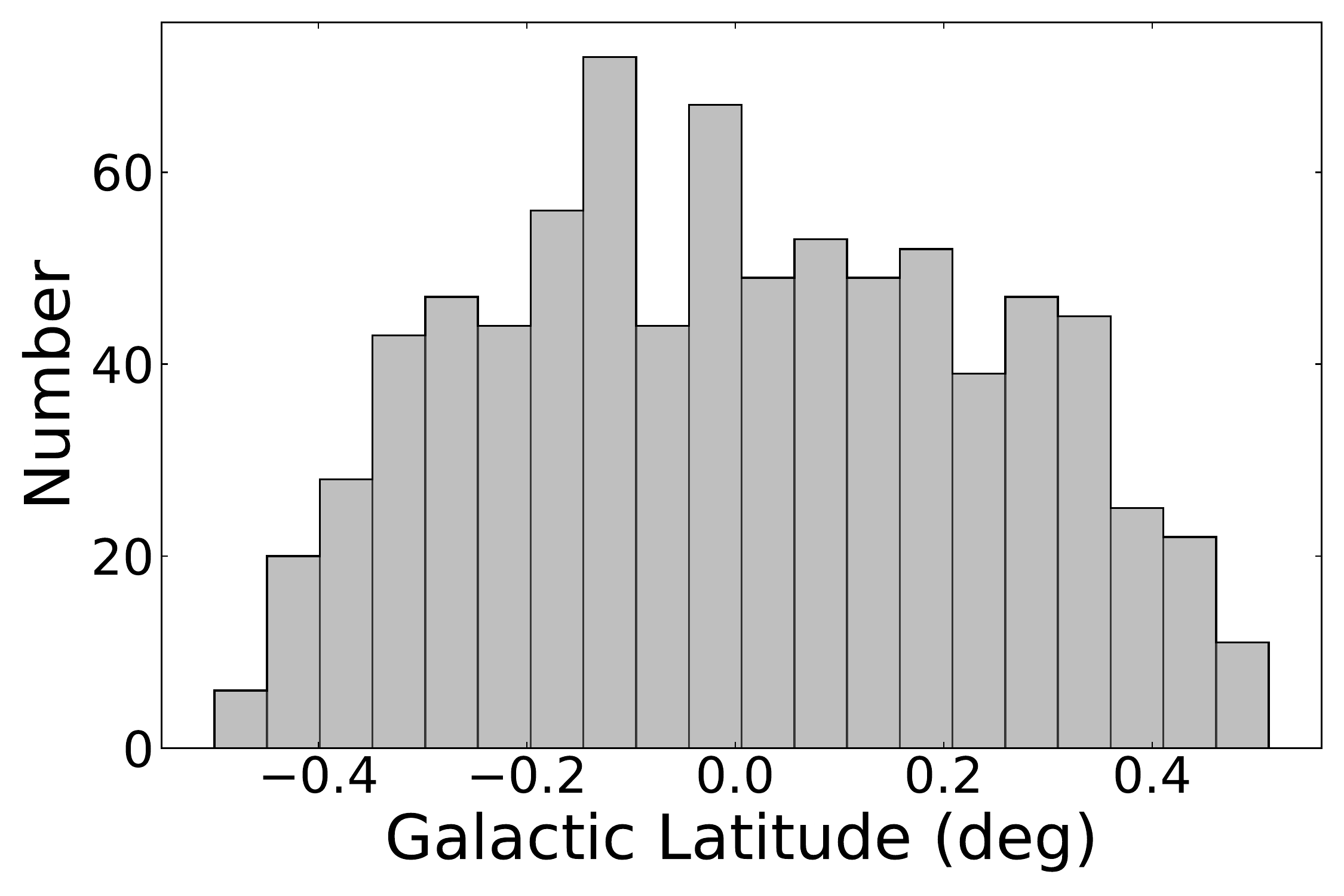}
\caption{Histograms of the number of sources detected per Galactic longitude on both sides of the plane and in Galactic latitude (combining both sides of the plane). We note that the dip in sources at $l$\,$\approx$\,$14$ deg and $340$ deg are due to stray light contaminating those fields. 
}
\label{fig: hist_lat_lon}
\end{figure}

 \begin{figure*} 
\centering
\includegraphics[width=2\columnwidth]{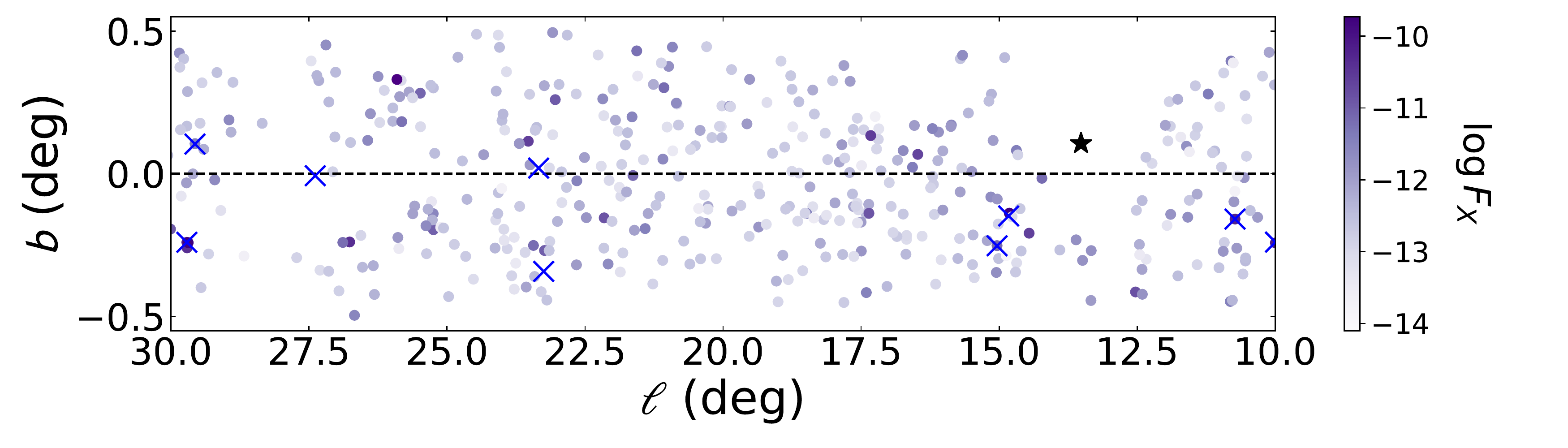}
\includegraphics[width=2\columnwidth]{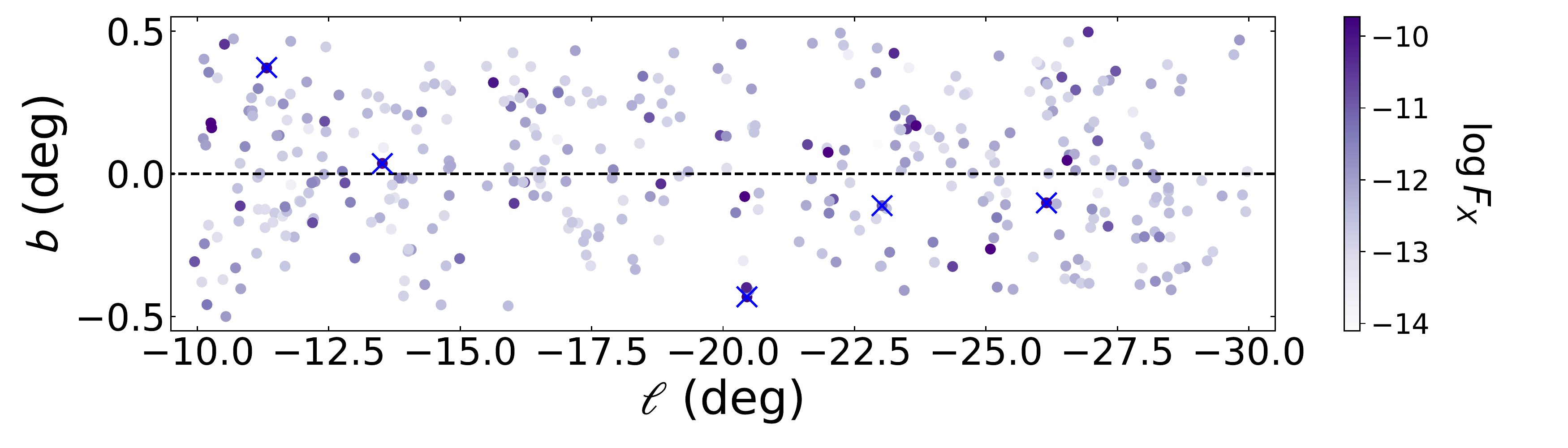}
\caption{The location of DGPS sources in Galactic coordinates. The sources are colored based on the logarithm of their X-ray flux. The blue crosses show the locations of known magnetars. The black star (bottom panel) marks a dominant source of stray light, leading to an obvious lack of sources at that region of the Survey. 
}
\label{fig: l_vs_b}
\end{figure*}

\subsection{New or Newly Classified Sources}
\label{sec: newly classified}

%Do we report our ToO observations here like SBS did. Need a short section on all the sources we wrote papers on...

We followed up unclassified, variable sources using our approved ToOs on \textit{Chandra}, \textit{NuSTAR}, \textit{NICER}, and \textit{XMM-Newton}, including  \textit{XMM-Newton} AO17 (Proposal ID: 082186; PI: Kouveliotou), \textit{Chandra} Cycles 19, 20, and 23 (Proposal IDs: 19500723, 20500298, and 23500070; PI: Kouveliotou), and \textit{NICER} Cycle 3 and 4 (Proposal IDs: 4050 and 5097; PI: Kouveliotou). 
The DGPS was a \textit{NuSTAR} Legacy Survey until 2019, although since this time we have utilized Director's Discretionary Time (DDT) observations. 
In total, we carried out 3 \textit{XMM-Newton} ToOs, 9 \textit{NuSTAR} ToOs, 9 \textit{Chandra} ToOs, 6 \textit{NICER} ToOs, and 20 \textit{Swift} ToOs to follow DGPS sources. 
%Values as of March 1, 2022
%We also successfully requested 19 \textit{Swift} ToOs %(PI: O'Connor) 
%of DGPS sources. 
%Values as of March 1, 2022
In addition, we made use of multi-wavelength observations from the Lowell Discovery Telescope (LDT), the South African Astronomical Observatory (SAAO) 1-m telescope, and the Southern African Large Telescope (SALT). The results of these campaigns were reported in \citet{Gorgone2019,Gorgone2021,OConnor2021,OConnor2023polar,OConnor2023IP}.

\subsection{Machine Learning Classification of DGPS Sources}
\label{sec: machinelearning}

As shown in \S \ref{sec: classification}, the DGPS has detected a large number of unclassified X-ray sources. The classification of hundreds of X-ray sources based on manual compilation and analyses of multi-wavelength datasets is difficult and time consuming. Instead, it is more efficient to turn to supervised machine learning methods to perform the classification of a large number of sources based on the properties of a training dataset comprising sources with already known classes. \citet{Yang2022} performed such analysis for a subset of the \textit{Chandra} Source Catalog version 2.0 (CSCv2) using a publicly available\footnote{\url{https://github.com/huiyang-astro/MUWCLASS_CSCv2}} \texttt{Python} framework and a training dataset of $\sim$3,000 sources with verified classifications\footnote{\url{https://home.gwu.edu/ ~kargaltsev/XCLASS/}}. They first applied a selection criterion to CSCv2 to remove \textit{Chandra} sources with low signal-to-noise, poor localization errors, or those that were either extended or confused (see \citealt{Yang2022} for details). The sources satisfying their criteria are referred to as ``good'' CSCv2 sources (GCS). In total, they are able to provide classifications to 66,359 CSCv2 sources, approximately 21\% of the CSCv2 catalog. 

While \citet{Yang2022} have not yet extended their analysis to other X-ray missions (see, however, \citealt{Tranin2022}), their results can provide useful insight into the classification of a subset of DGPS sources. We note that one of the main obstacles for extending these analyses to \textit{Swift} is the significantly larger localization uncertainties of X-ray sources precluding accurate multi-wavelength cross-matching. Therefore, below we only review the classifications of DGPS sources that have counterparts in CSCv2, which provide much more accurate positions.  
%regions and the lack of confidence in the multi-wavelength counterpart, which can significantly impact the source classification. 

\begin{figure}[h]
\centering
\includegraphics[width=\columnwidth]{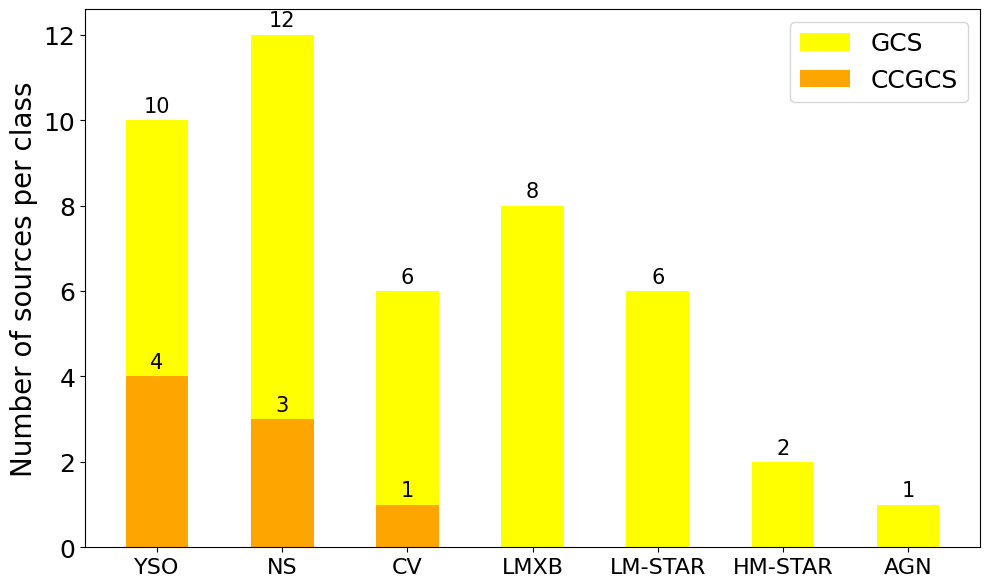}
\caption{Histogram of source classification breakdown for 45 GCSs and 8 CCGCSs based on results from \citet{Yang2022}.
} 
\label{fig: classification}
\end{figure}

After performing a cross-match between DGPS sources and the CSCv2 catalog we find 186 matches (Table \ref{tab: cross-match}). We then matched these sources to the results of \citet{Yang2022}, finding 45 classified GCSs in addition to 19 sources in their training datset. These sources have a classification confidence threshold (CT) indicating the confidence level, with  CT$\geq$2 adopted for confidently classified GCSs (CCGCSs). Out of the 45 GCS sources, only 8 are CCGCSs. In Figure \ref{fig: classification} we display the classification stacked histogram of all 45 sources. The largest number of CCGCSs are 4 YSOs, followed by 3 NSs, and 1 CV.

Although 3 NS candidates (2CXO J171428.6--383601, 2CXO J182524.7--114524 and 2CXO J181210.3--184208), which each lack any optical or infrared counterpart, have been confidently classified, this may be due to a bias in the training dataset against faint sources without multi-wavelength counterparts. A large fraction of faint sources do not have multi-wavelength counterparts simply because of the insufficient sensitivity of optical and infrared surveys combined with the significant extinction in the GP. The classification algorithm of \citet{Yang2022} may instead interpret the lack of multi-wavelength counterparts as a sign of the NS class (which includes both magnetars and isolated NSs). 
Indeed, upon further investigation, 2 out of 3 of these NS candidates (2CXO J182524.7--114524 and 2CXO J181210.3--184208) have infrared counterparts in UKIDSS, which is significantly more sensitive than the 2MASS catalog used in \citet{Yang2022}. The third source (2CXO J171428.6--383601) may have an infrared counterpart in VVV, but the source lies outside of the 95\% localization region (0.9\arcsec) from CSCv2 at an offset of 1.2\arcsec. Based on the VVV sky density in this region of the GP, we compute a probability of chance coincidence of between $25-37\%$, depending on whether or not we account for the brightness of the counterpart.

%the results of the selection bias that the GCS sources are systematically fainter than the sources from the training dataset, and may be preferably classified as NSs due to their faintness and lacking of multiwavelength counterparts \citep{Yang2022}. 
%\brendan{we can remove this struck out paragraph?}{\textco;or{blue} {\bf CK: YSOs can be anything (including a magnetar) so yes, this is not very informative and I agree to delete it}

\subsection{Constraints on the population of magnetars}

The main targets of the \textit{Swift} DGPS were magnetars and HMXBs. However, although several of the already known sources from both populations were observed (Figure \ref{fig: class_pie}), we did not concretely identify any new transient events associated with magnetars, and classified only a single new HMXB \citep{OConnor2021}. 

Magnetars are generally identified during their bright X-ray outbursts. As such, the quiescent magnetar population is poorly constrained. Using the Magnetar Outburst Online Catalog\footnote{\url{http://magnetars.ice.csic.es/\#/welcome}} \citep{CotiZelati2018}, we compiled the distance and quiescent X-ray ($0.3$\,$-$\,$10$ keV) luminosity for 15 magnetars. Their observed quiescent luminosities lie between $10^{30-35}$ erg s$^{-1}$ \citep{CotiZelati2018}. Using the best available distance for each event, we find quiescent X-ray fluxes in the range $10^{-15}$ to $10^{-12}$ erg cm$^{-2}$ s$^{-1}$. Therefore, only 7 out of 15 magnetars would be detectable based on the DGPS 50\% completeness flux. 

For example, we note here that the DGPS observed the field of the magnetar \textit{Swift} J\,$1818.0-1607$ \citep{Blumer2020,Champion2020,ChinPing2020} approximately $2.7$ yr before its discovery. Unfortunately the source was not active and we were only able to obtain an upper limit ($3\sigma$) of $\lesssim 2\times10^{-13}$ erg cm$^{-2}$ s$^{-1}$. 
This demonstrates that quiescent magnetars exist in the region covered by the DGPS, but their identification is difficult, possibly due to faintness. A significant benefit of this survey is to constrain the quiescent luminosity of future magnetars, or other transients, discovered in these regions.

In fact, \citet{Beniamini2019magnetar} 
found that based on the observed persistent luminosity and $\log N$\,$-$\,$\log S$ distribution, the number of hidden magnetars could outweigh the known population by a factor of up to $\sim$10. They found that the missing magnetars should have unabsorbed fluxes $<$\,$10^{-13}$ erg cm$^{-2}$ s$^{-1}$, which is below the DGPS completeness values. 

In the general spin-down model for magnetars the magnetic field evolution is parameterized by $\dot{B}\propto B^{1+\alpha}$ \citep{Colpi2000}. \citet{Beniamini2019magnetar} used the observed $\log N$\,$-$\,$\log S$ for magnetars to show that both $\alpha$\,$=$\,$0$ and $-1$ can explain the observed population of absorbed and unabsorbed magnetar fluxes. We perform a similar calculation using the constraints of our Survey. Based on the DGPS $\log N$\,$-$\,$\log S$ (Figure \ref{fig: logN-logS}) we have detected 144 sources at $>$\,$1.0\times 10^{-12}$ erg cm$^{-2}$ s$^{-1}$ of which 10 are known magnetars (Figure \ref{fig: class_pie}) and 400 sources at $>$\,$2.7\times 10^{-13}$ erg cm$^{-2}$ s$^{-1}$ (including the 144 mentioned above). Under the assumption that none of these new sources are magnetars we constrain $\alpha$\,$<$\,$-0.65$ at the 90\% confidence level (CL). We note that the assumption that none of the $\sim$1,000 sources in our Survey are magnetars is likely too restrictive as there could be unidentified quiescent magnetars hiding in this population. If instead we assume there are 10 (20) unidentified magnetars with a flux between $2.7\times 10^{-13}$ to $1.0\times 10^{-12}$ erg cm$^{-2}$ s$^{-1}$ the constraint is $\alpha$\,$<$\,$0.86$ ($2.15$). These results are consistent with \citet{Beniamini2019magnetar}.

The upper limit to $\alpha$ is therefore strongly dependent on the unknown population of unidentified quiescent magnetars hiding in our sample. Nevertheless, the identification of their quiescent population is extremely difficult. This issue was explored in detail by \citet{Muno2008} using constraints from \textit{XMM-Newton} and \textit{Chandra}. They searched for periodic variability in deep X-ray observations of the GP region ($|b|$\,$<$\,$5$ deg), but did not identify any new periods between 5 and 20 s. Based on their analysis, \citet{Muno2008} found that $<540$ magnetars (90\% CL) should exist in the Milky Way. Due to the lower exposure times and photon counts of our Survey compared to the deep \textit{XMM-Newton} and \textit{Chandra} data used by \citet{Muno2008}, a timing analysis of our sources is not as fruitful. 

%\citet{CotiZelati2018} demonstrated an anti-correlation between the increase in X-ray luminosity and quiescent X-ray luminosity, such that $R_{flux}\approx\Delta L_X=L_{X,\textrm{peak}}/L_{X,\textrm{quiescent}}\propto L_{X,\textrm{quiescent}}^{-0.7}$. This conveys that magnetars with lower persistent luminosities ($10^{31-33}$ erg s$^{-1}$ experience the largest outbursts. 

%Magnetar outbursts have been shown to follow a power-law energy distribution $dN/dE$\,$\propto$\,$ E^{-\gamma}$ where $\gamma$\,$=$\,$1.4$\,$-$\,$1.8$ \citep{Gogus1999,Gogus2000}. 

\section{Conclusions}
\label{sec: conclusions}

We have presented the results of the DGPS Phase-I observations, covering Galactic longitude $10$\,$<$\,$|l|$\,$<$\,$30$ deg and latitude $|b|$\,$<$\,$0.5$ deg. These observations led to the identification of 928 %802+126
unique X-ray sources (Tables \ref{tab: main_cat} and \ref{tab: not_in_LSXPS}) of which 358 (40\%) %249+109
were previously unknown to other X-ray surveys. Our results indicate a significant population of very faint X-ray sources below $F_X$\,$<$\,$10^{-13}$ erg cm$^{-2}$ s$^{-1}$, emphasizing the necessity for sensitive, next generation, wide-field X-ray telescopes (e.g., \textit{Athena}, \citealt{Athena}; \textit{AXIS}, \citealt{Axis2019}; \textit{Lynx}, \citealt{Lynx2019}; \textit{STAR-X}, \citealt{starx}) to characterize the missing faint X-ray population in our Galaxy.

\acknowledgments

The authors thank the anonymous referee for constructive feedback that improved the manuscript. 
B.~O. and C.~K. acknowledge support from multiple grants to follow-up DGPS sources: NASA Grants 80NSSC17K0335, 80NSSC20K0389, 80NSSC19K0916, 80NSSC22K1398, and 80NSSC22K0583, \textit{Chandra} awards GO9-20057X and GO2-23038X, and award number 46939-1-CCNS 22317F. 
P.~B.'s research was supported by a grant (no. 2020747) from the United States-Israel Binational Science Foundation (BSF), Jerusalem, Israel. J.~H. and Z.~W. acknowledge support from NASA under award number 80GSFC21M0002.

This work made use of data supplied by the UK \textit{Swift} Science Data Centre at the University of Leicester. This research has made use of the XRT Data Analysis Software (XRTDAS) developed under the responsibility of the ASI Science Data Center (ASDC), Italy. This research has made use of the VizieR catalogue access tool, CDS, Strasbourg, France (DOI: 10.26093/cds/vizier). The original description of the VizieR service was published in A\&AS 143, 23. This research has made use of the SIMBAD database, operated at CDS, Strasbourg, France. This research has made use of data obtained from the Chandra Source Catalog, provided by the Chandra X-ray Center (CXC) as part of the Chandra Data Archive. This research has made use of data obtained from the 4XMM XMM-Newton serendipitous source catalogue compiled by the 10 institutes of the XMM-Newton Survey Science Centre selected by ESA. This research has made use of data and/or software provided by the High Energy Astrophysics Science Archive Research Center (HEASARC), which is a service of the Astrophysics Science Division at NASA/GSFC.

%This research has made use of the NuSTAR Data Analysis Software (NuSTARDAS) jointly developed by the ASI Space Science Data Center (SSDC, Italy) and the California Institute of Technology (Caltech, USA).
    %acknowledge all instruments used properly!!


\vspace{5mm}
\facilities{\textit{Swift}/XRT
}


\software{\texttt{HEASoft}, \texttt{XRTDAS}, \texttt{swifttools} \citep{LSXPS}, \texttt{PIMMS}, Astropy \citep{astropy}}



\newpage
\appendix

\section{Additional GP Mosaics}
\label{sec: appendix_mosaics}

\begin{figure*} 
\centering
\includegraphics[width=2.15\columnwidth]{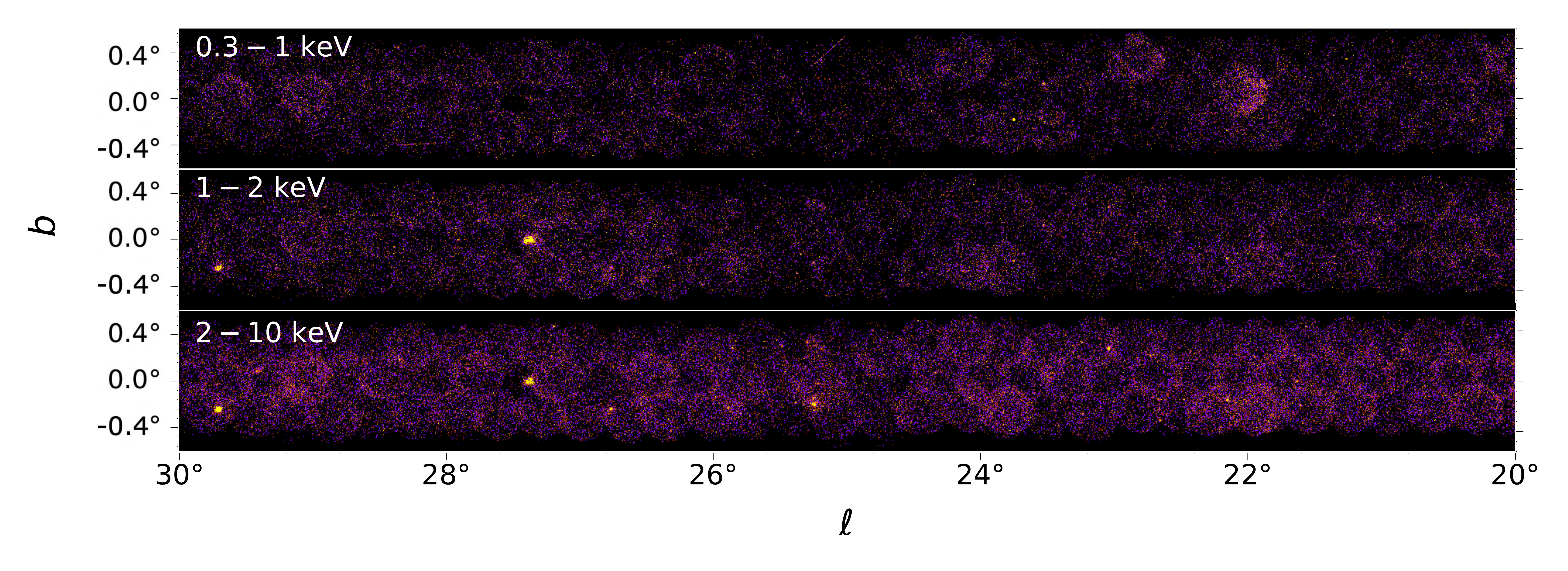}
\includegraphics[width=2.15\columnwidth]{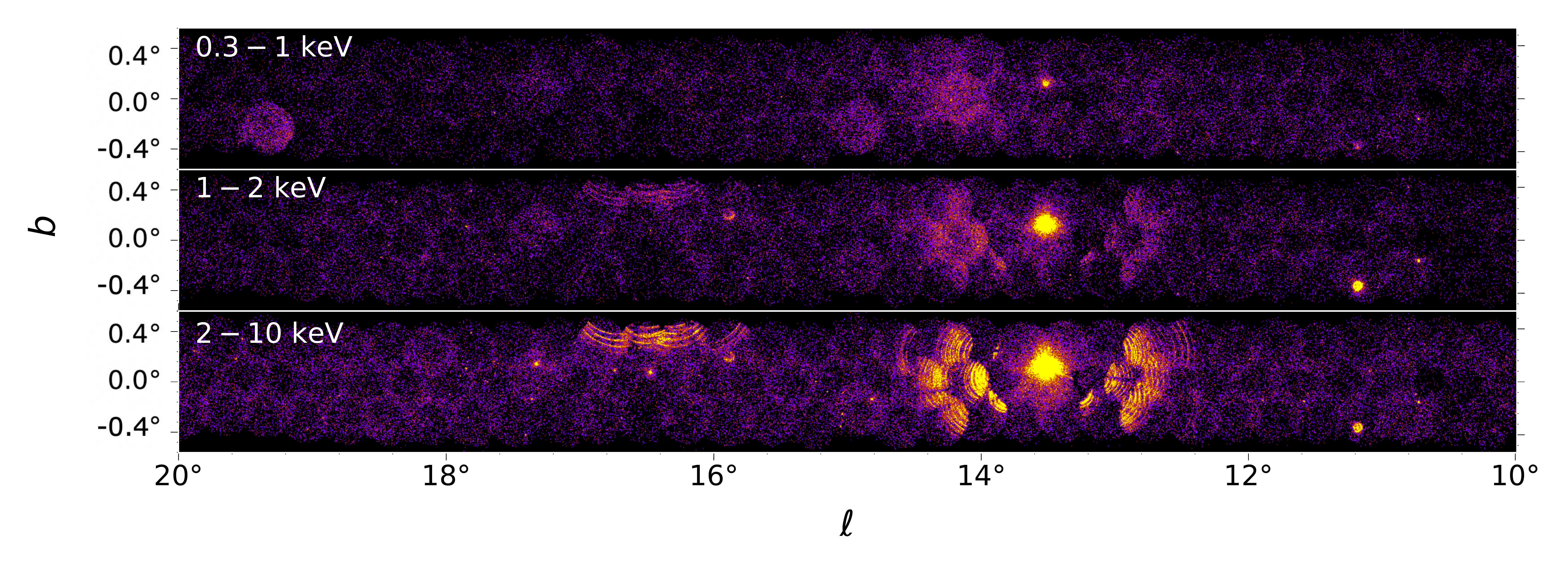}
\caption{Mosaic of the GP at positive galactic longitudes in the SB ($0.3$\,$-$\,$1$ keV), MB ($1$\,$-$\,$2$ keV), and HB ($2$\,$-$\,$10$ keV).
}
\label{fig: GP_mosaic_3bands_right}
\end{figure*}

\begin{figure*} 
\centering
\includegraphics[width=2.15\columnwidth]{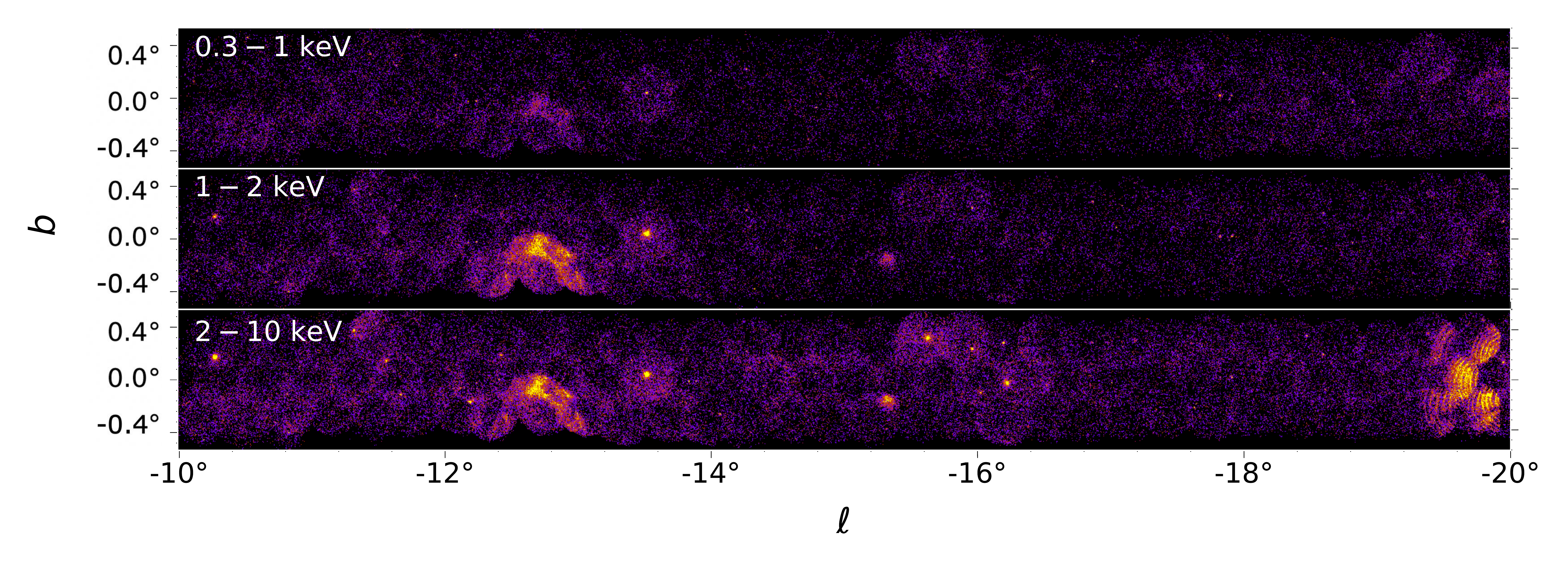}
\includegraphics[width=2.15\columnwidth]{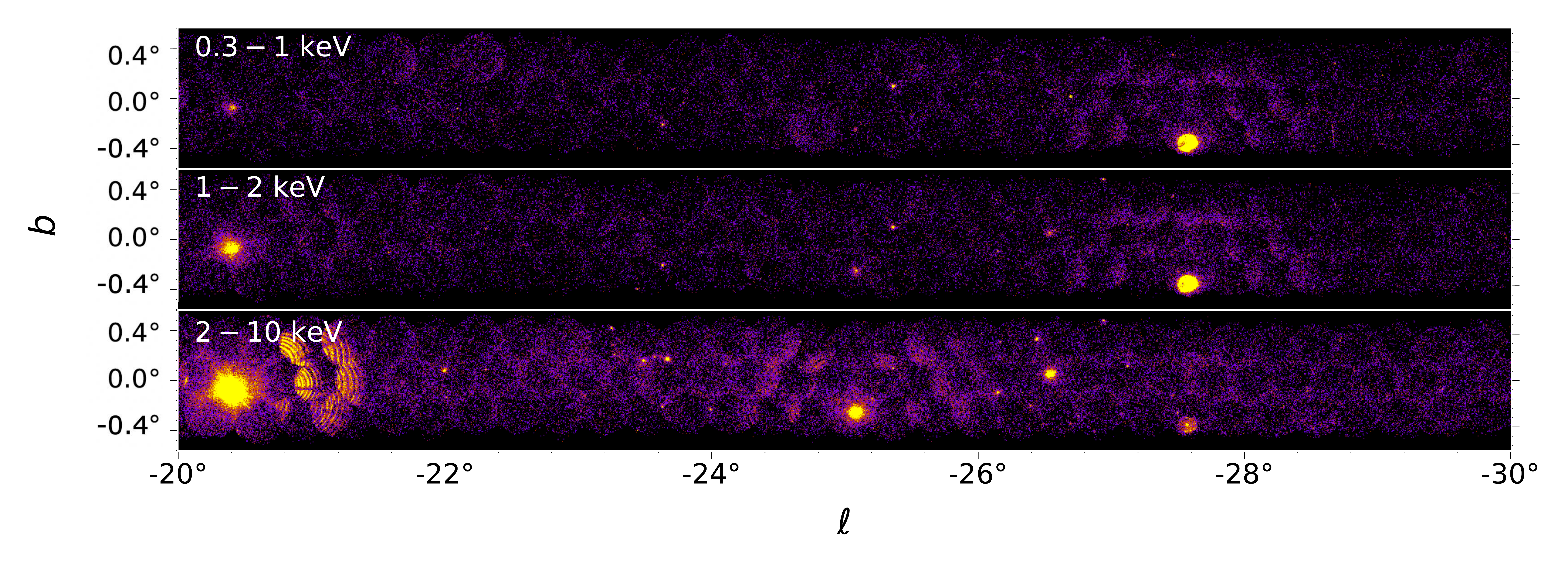}
\caption{Mosaic of the GP at negative galactic longitudes in the SB ($0.3$\,$-$\,$1$ keV), MB ($1$\,$-$\,$2$ keV), and HB ($2$\,$-$\,$10$ keV). The background variation acrsoss the plane is due to higher exposure in regions with overlapping tiles and is not due to an intrinsic structure in the emission. 
}
\label{fig: GP_mosaic_3bands_left}
\end{figure*}

Here we present additional mosaics of the DGPS observations in the SB, MB, and HB (Figures \ref{fig: GP_mosaic_3bands_left} and \ref{fig: GP_mosaic_3bands_right}). These mosaics complement the FB image of the plane displayed in Figure \ref{fig: GP_mosaic}.

\section{Comparison of source properties in Galactic coordinates}
\label{sec: appendix_galactic_coord_figs}

\begin{figure*} 
\centering
\includegraphics[width=2\columnwidth]{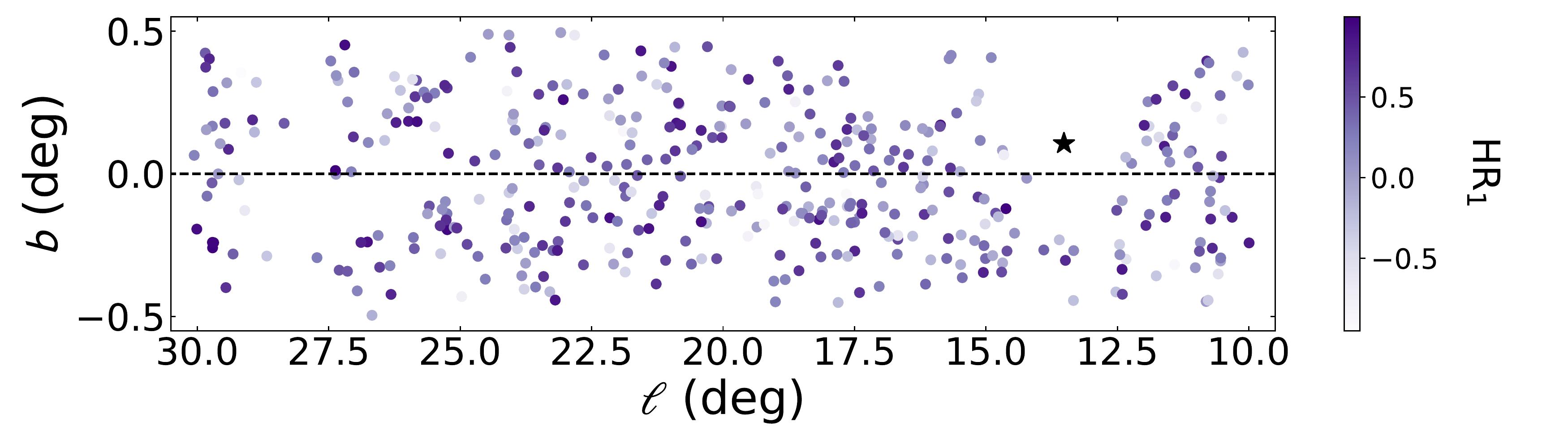}
\includegraphics[width=2\columnwidth]{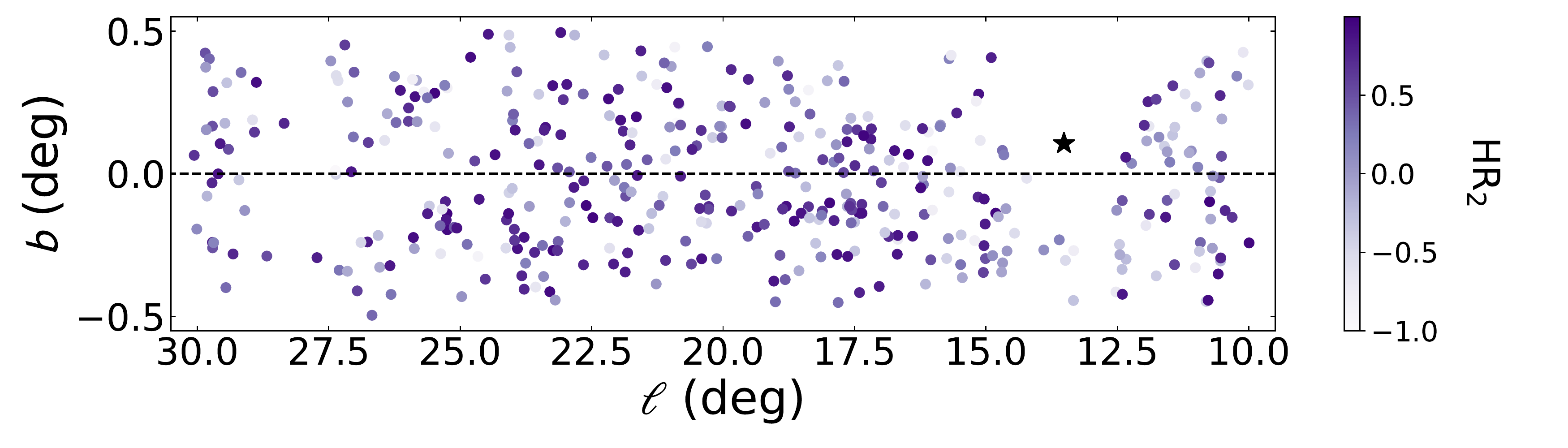}
\includegraphics[width=2\columnwidth]{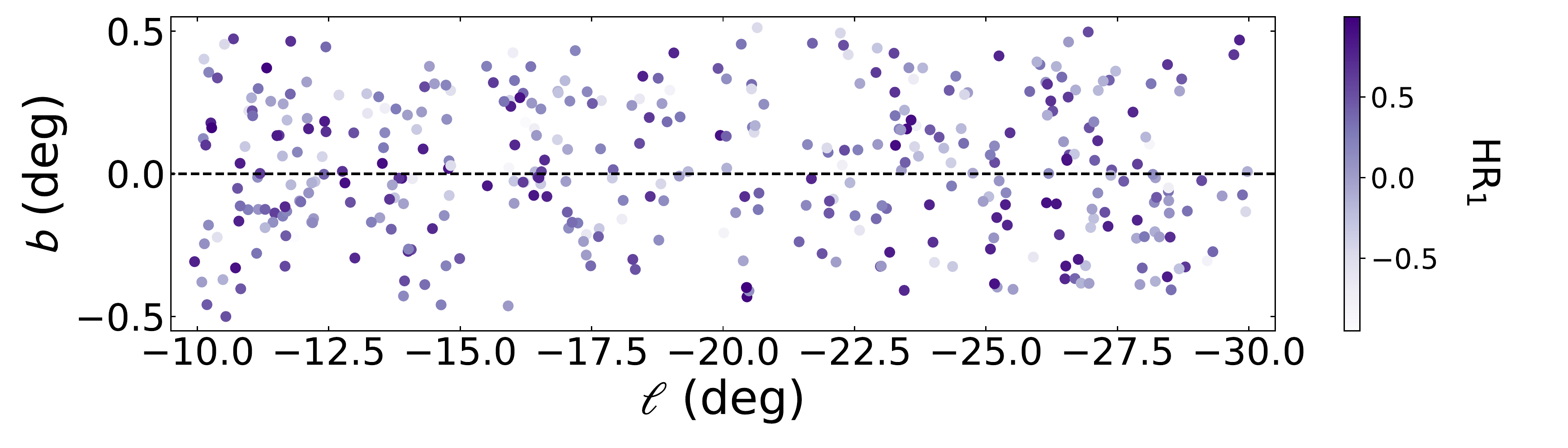}
\includegraphics[width=2\columnwidth]{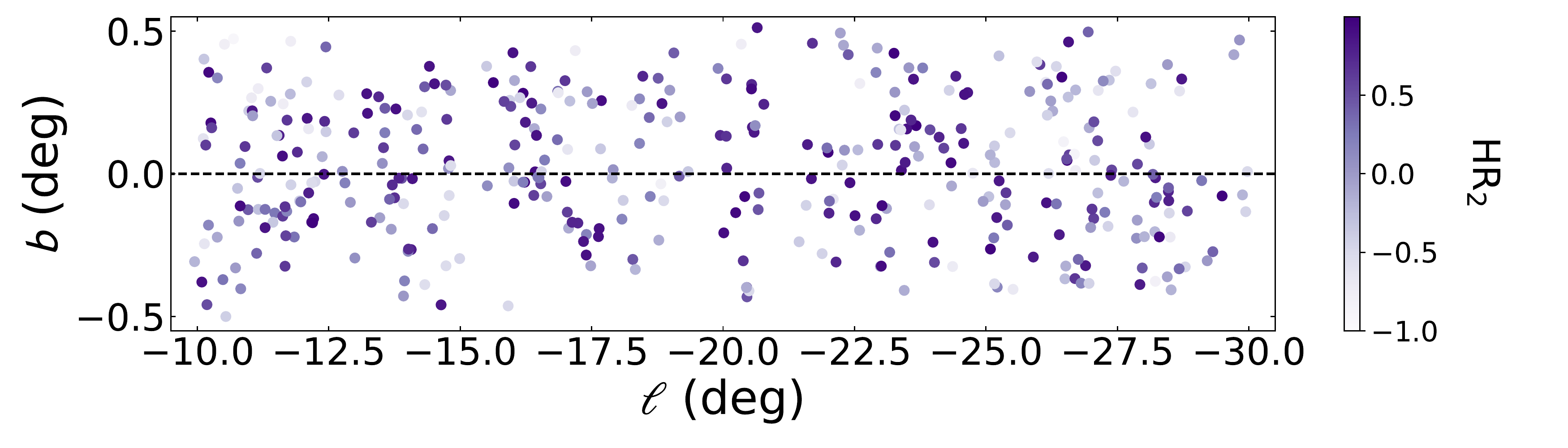}
\caption{The location of DGPS sources in Galactic coordinates. The sources are colored based on their hardness ratio (either $HR_1$ or $HR_2$). The black star (bottom panels) marks a dominant source of stray light, leading to an obvious lack of sources in that region of the Survey.
}
\label{fig: l_vs_b_HR}
\end{figure*}

Here we present additional figures demonstrating how source properties vary with location in the Galactic Plane. Figure \ref{fig: l_vs_b_HR} shows the hardness ratio for each source versus their location in Galactic coordinates. There appears to be a clustering of sources in $HR_2$, but less so in $HR_1$. We note that the hardness ratios are uncorrected for Galactic hydrogen column density, and that a line of sight absorption effect may be at play here.

\begin{figure*} 
\centering
\includegraphics[width=\columnwidth]{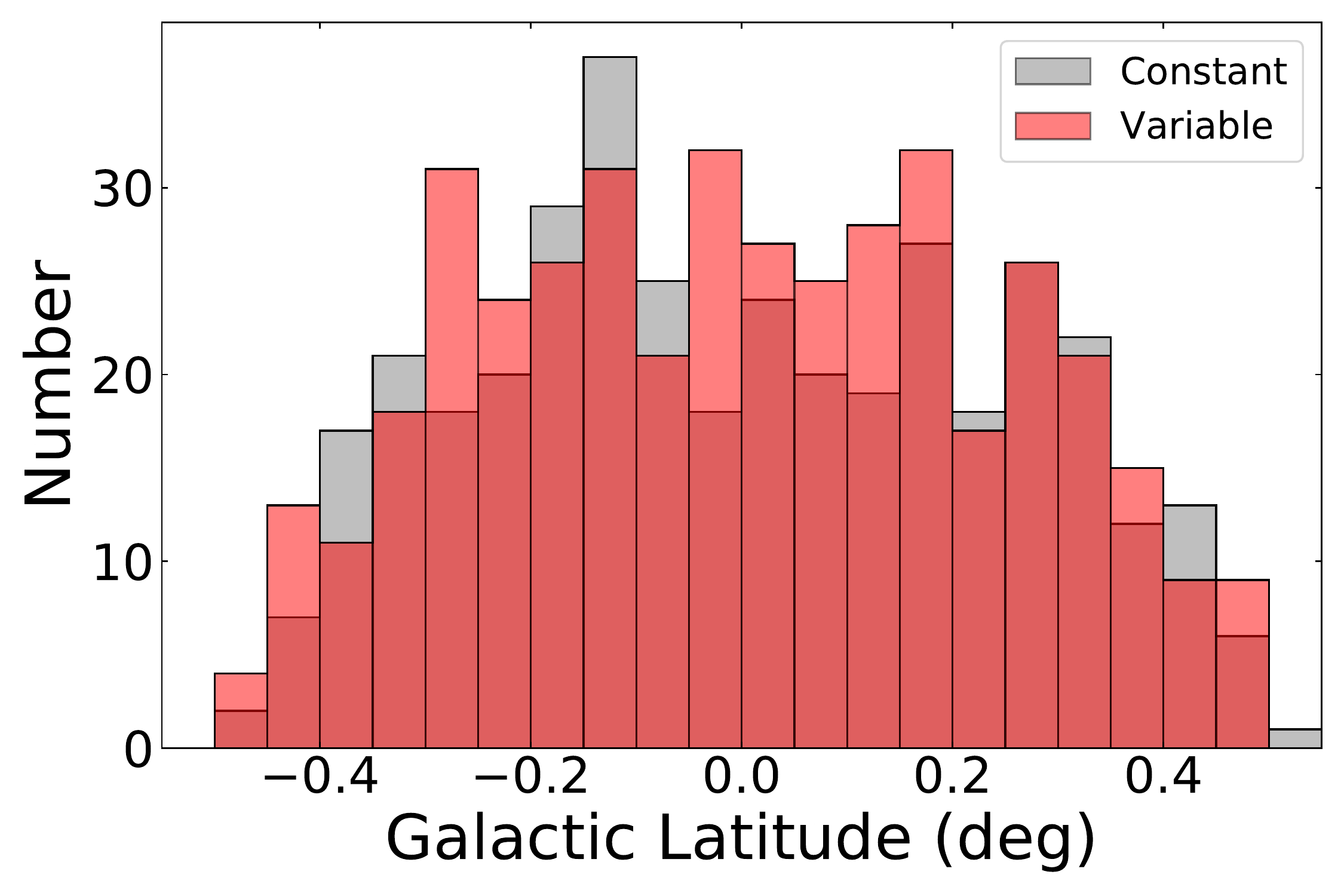}
\includegraphics[width=\columnwidth]{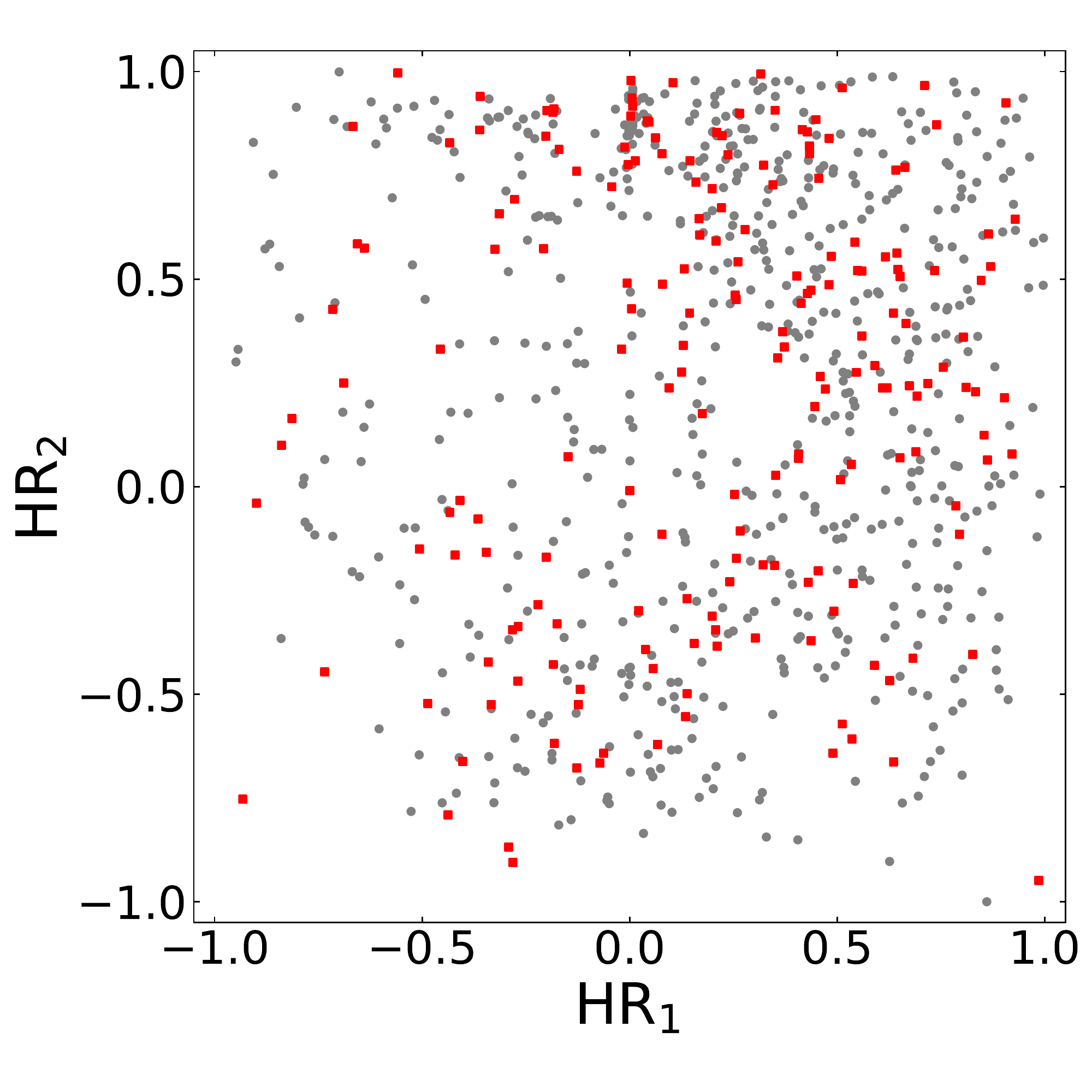}
\caption{\textbf{Left:} Histogram of source location in Galactic latitude for constant (gray) and variable (red) sources (see \S \ref{sec: variable}). 
\textbf{Right:} Distribution of sources in the hardness ratio plane. Red squares are sources on the GP with $|b|$\,$<$\,$0.1$ deg and gray circles are off-plane sources with $|b|$\,$>$\,$0.1$ deg
}
\label{fig: b_var_hist}
\end{figure*}

In Figure \ref{fig: b_var_hist} (left) we show a histogram of Galactic latitude for variable and constant sources. There is no discernible difference and a Kolmogorov-Smirnov test supports the null hypothesis ($p-$value\,$=$\,$0.7$) that they are drawn from the same distribution.

We also show the source distribution in the hardness ratio plane separated between $|b|$\,$<$\,$0.1$ deg and $|b|$\,$>$\,$0.1$ deg (Figure \ref{fig: b_var_hist}; right). There is no obvious clustering of sources based on this separation criterion.

\section{Derivation of Hardness Ratios for X-ray Source Populations}
\label{sec: HR derivation}

The majority of sources detected with the DGPS are faint, with a low number of source counts (i.e., $<30$ cts), and, therefore, an analysis of their X-ray spectra does not provide strong constraints on the intrinsic source properties. Therefore, we utilized the X-ray hardness ratios, comparing the count rate between different energy bands, as a way to characterize source spectra despite the small number of counts. The hardness ratios $HR_1$ and $HR_2$ are defined as in \citet{Evans2014,Evans2020}:
\begin{equation}
    HR_1 = \frac{MB-SB}{MB+SB}\,\, \textrm{and}\,\,
    HR_2 = \frac{HB-MB}{HB+MB},
\end{equation}
where $SB=0.3-1$ keV, $MB=1-2$ keV, and $HB=2-10$ keV count rate. The use of two hardness ratios is ideal for characterizing soft sources, and distinguishing between different source classifications. %We note that a third hardness ratio can be developed between the HB and the SB as $HR_3 = $ \textcolor{red}{HR2-HR1 does not reduce to HB-SB/HB+SB ... so nevermind}

%In order to characterize the expected location of magnetars in the $HR_1$-$HR_2$ plane, we utilized the McGill Online Magnetar Catalog\footnote{\url{https://www.physics.mcgill.ca/~pulsar/magnetar/main.html}} to compile the spectral properties of 30 known magnetars as of 26 August 2022. The majority of magnetar spectra are compiled while the magnetar is in outburst, leading to its discovery. However, we utilize the persistent (quiescent) magnetar properties in order to explore their expected hardness ratios.  %Magnetar spectra fall into a few categories: BB, 2BB, 3BB, BB+PL, 2BB+PL etc... (see Coti Zelati and cite as well - ask george if any of his papers need citing here) The majority of magnetar spectra are compiled while the magnetar is in outburst, leading to its discovery. However, the quiescent magnetar population is poorly constrained. For example, J1818 was in the DGPS footprint but non-detected prior to its outburst in March 2020 etc etc. Quiescent luminosity between $10^{30-35}$ \citep{CotiZelati2018} so detectable by survey out to XX kpc using the DGPS 50\% completeness flux.Magnetars at $<5$ kpc mainly BB or PL+BB so we focused on these populations...

In order to characterize the expected location of different source classes in the $HR_1$\,$-$\,$HR_2$ plane we assumed spectral properties belonging to each class and varied the hydrogen column density \citep[see also][]{Rigoselli2022}. We did this for HMXBs assuming a power-law spectrum with photon index $\Gamma$\,$=$\,$1$, for stars assuming an \texttt{APEC} spectrum with temperature $kT$\,$=$\,$1.085$ keV and 0.6 solar abundance, and for magnetars assuming a blackbody with $kT$\,$=$\,$1$ keV. We varied the hydrogen column density uniformly between $\log(N_H/\textrm{cm}^{-2})$\,$=$\,$18$\,$-$\,$23$. 
We performed this calculation using \texttt{PIMMS} to compute the \textit{Swift}/XRT count rate in the SB, MB, and HB at each step in the grid. We then determined both hardness ratios based on these values. We show the tracks of each source type in Figure \ref{fig: HR-plane}. The majority of stars have thermal plasma temperatures less than $kT$\,$<$\,$1$ keV, such that they lie below the line in $HR_1$\,$-$\,$HR_2$ space. Similarly, many HMXBs display harder spectra than $\Gamma$\,$=$\,$1$, and for that reason lie above the line in $HR_1$\,$-$\,$HR_2$ space. In the case of magnetars, their quiescent spectra are generally described by a softer blackbody with $kT$\,$\approx$\,$0.4$ keV \citep{CotiZelati2017}, suggesting that quiescent magnetars will lie below the line drawn. 

We further checked the observed location of different source classes in the $HR_1$\,$-$\,$HR_2$ plane by obtaining the observed mean flux and mean hardness ratios from the 2SXPS catalog. In Figure \ref{fig: HR-plane} (right) we show the observed locations for magnetars from the McGill Online Magnetar Catalog \citep{Olausen2014}, HMXBs from \citet{Liu2006}, LMXBs from \citet{Liu2007}, and IP CVs from Koji Mukai's online catalog. As expected, many HMXBs lie above our computed line for $\Gamma$\,$=$\,$1$. Of further note is the broad diversity observed for magnetars, possibly due to the observed outbursts by \textit{Swift} and the spectral cooling of the sources during outburst \citep{CotiZelati2018}.

\section{Tables of catalog contents}

Here we provide a description of the contents available for the DGPS catalog:
\begin{enumerate}
    \item Sources with additional information pulled from LSXPS (Table \ref{tab: main_cat})
    \item Sources not in LSXPS (Table \ref{tab: not_in_LSXPS})
\end{enumerate}
The main difference between the catalogs is the availability of variability and hardness ratio information. Both catalogs comprise the full result of the DGPS Phase-I.

In these tables, the energy bands are coordinated such that they agree with the \citet{LSXPS} definitions: 
band0 is the full band (FB; $0.3$\,$-$\,$1$ keV), band1 is the soft band (SB; $0.3$\,$-$\,$10$ keV), band2 is the medium band (MB; $1$\,$-$\,$2$ keV), band3 is the HB (HB; $2$\,$-$\,$10$ keV).

\begin{table*}
\centering
\caption{Contents for DGPS sources for which we were able to pull additional information from LSXPS. The table is accessible in electronic form through VizieR \citep{Ochsenbein2000}.}
\label{tab: main_cat}
\begin{tabular}{lcc}
\hline\hline
Column & Units & Description \\
\hline
IAU Name	& & IAU name in format ``DGPS JHHMMSS.S$\pm$DDMMSS'' \\
LSXPS\_ID	& &  Numerical unique source identifier within LSXPS \\
RA	& deg & Right Ascension (J2000) \\
DEC	& deg & Declination (J2000) \\
Err90		&arcsec & 90\% source position uncertainty \\
l	& deg & Galactic longitude \\
b	& deg & Galactic latitude \\
Rate\_band0	& cts s$^{-1}$ & FB ($0.3$\,$-$\,$10$ keV) count rate \\
Rate\_band0\_pos	& cts s$^{-1}$& Positive count rate error \\
Rate\_band0\_neg	&cts s$^{-1}$ & Negative count rate error \\
Rate\_band1	&cts s$^{-1}$ & SB ($0.3$\,$-$\,$1$ keV) count rate \\
Rate\_band1\_pos	& cts s$^{-1}$&Positive count rate error \\
Rate\_band1\_neg	& cts s$^{-1}$&  Negative count rate error\\
Rate\_band2	&cts s$^{-1}$ & MB ($1$\,$-$\,$2$ keV) count rate \\
Rate\_band2\_pos	&cts s$^{-1}$ &Positive count rate error \\
Rate\_band2\_neg	&cts s$^{-1}$ & Negative count rate error \\
Rate\_band3	&cts s$^{-1}$ & HB ($2$\,$-$\,$10$ keV) count rate \\
Rate\_band3\_pos	&cts s$^{-1}$ &Positive count rate error \\
Rate\_band3\_neg	&cts s$^{-1}$ & Negative count rate error \\
FixedPowUnabsFlux	& erg cm$^{-2}$ s$^{-1}$ & FB X-ray flux ($0.3$\,$-$\,$10$ keV) assuming a photon index $\Gamma$\,$=$\,$1.7$ \\
FixedPowUnabsFlux\_pos	&erg cm$^{-2}$ s$^{-1}$ & Positive flux error \\
FixedPowUnabsFlux\_neg	&erg cm$^{-2}$ s$^{-1}$ & Negative flux error \\
R\_Flux & & Ratio of peak-to-mean flux \\
HR1	& & Hardness ratio between the MB and SB \\
HR1\_pos	& & Positive error on hardness ratio\\
HR1\_neg	& & Negative error of hardness ratio \\
HR2	& & Hardness ratio between the HB and MB \\
HR2\_pos	& & Positive error on hardness ratio \\
HR2\_neg	& & Negative error of hardness ratio \\
GalacticNH	& cm$^{-2}$ & Hydrogen column density in the source direction \citep{Willingale2013} \\
Exposure	& s & Cumulative DGPS exposure at the source position\\
X-ray Match	& & ``Y'' if known X-ray source, otherwise ``N'' \\
%Classified	& & ``Y'' if known classified source, otherwise ``N'' \\
Variable	& & ``Y'' if known source is variable, otherwise ``N'' \\
\hline\hline
\end{tabular}
\end{table*}

\begin{table*}
\centering
\caption{Contents for the non-LSXPS sources. These are sources with no LSXPS counterpart. The table is accessible in electronic form through VizieR \citep{Ochsenbein2000}.
}
\label{tab: not_in_LSXPS}
\begin{tabular}{lcc}
\hline\hline
Column & Units & Description \\
\hline
IAU Name	& & IAU name in format ``DGPS JHHMMSS.S$\pm$DDMMSS'' \\
RA	& deg & Right Ascension (J2000) \\
DEC	& deg & Declination (J2000) \\
Err90		&arcsec & 90\% source position uncertainty\\
X-ray Match	& & ``Y'' if known X-ray source, otherwise ``N'' \\
band0\_KNB91\_Detected & & Source retrospectively detected in FB using LSXPS Upper Limit Server  \\
& &(0 = not detected; 1 = detected) \\
band3\_KNB91\_Detected & & Source retrospectively detected in HB using LSXPS Upper Limit Server \\
band2\_KNB91\_Detected & &Source retrospectively detected in MB using LSXPS Upper Limit Server \\
band1\_KNB91\_Detected & &Source retrospectively detected in SB using LSXPS Upper Limit Server \\
band0\_IsDetected & & Source blindly detected in FB through iterative source detection on mosaics \\
& &(0 = not detected; 1 = detected) \\
band3\_IsDetected & &Source blindly detected in HB through iterative source detection on mosaics \\
band2\_IsDetected & &Source blindly detected in MB through iterative source detection on mosaics \\
band1\_IsDetected & &Source blindly detected in SB through iterative source detection on mosaics \\
Rate\_band0	& cts s$^{-1}$ & FB ($0.3$\,$-$\,$10$ keV) count rate \\
Rate\_band0\_pos	& cts s$^{-1}$& Positive count rate error \\
Rate\_band0\_neg	&cts s$^{-1}$ & Negative count rate error \\
Rate\_band3	&cts s$^{-1}$ & HB ($2$\,$-$\,$10$ keV) count rate \\
Rate\_band3\_pos	&cts s$^{-1}$ &Positive count rate error \\
Rate\_band3\_neg	&cts s$^{-1}$ & Negative count rate error \\
Rate\_band2	&cts s$^{-1}$ & MB ($1$\,$-$\,$2$ keV) count rate \\
Rate\_band2\_pos	&cts s$^{-1}$ &Positive count rate error \\
Rate\_band2\_neg	&cts s$^{-1}$ & Negative count rate error \\
Rate\_band1	&cts s$^{-1}$ & SB ($0.3$\,$-$\,$1$ keV) count rate \\
Rate\_band1\_pos	& cts s$^{-1}$&Positive count rate error \\
Rate\_band1\_neg	& cts s$^{-1}$&  Negative count rate error\\
FixedPowUnabsFlux	& erg cm$^{-2}$ s$^{-1}$ & FB X-ray flux ($0.3$\,$-$\,$10$ keV) assuming a photon index $\Gamma$\,$=$\,$1.7$ \\
FixedPowUnabsFlux\_pos	&erg cm$^{-2}$ s$^{-1}$ & Positive flux error \\
FixedPowUnabsFlux\_neg	&erg cm$^{-2}$ s$^{-1}$ & Negative flux error \\
\hline\hline
\end{tabular}
\end{table*}

%\begin{table*}
%\centering
%\caption{Sources not in LSXPS}
%\label{tab: not_in_LSXPS}
%\begin{tabular}{lcccccc}
%\hline\hline
% Name    & RA (J2000)  & Dec (J2000) & Err90 & Band & Rate & Flux \\
%\hline
%\hline
%& & & & \\
%& & & & \\
%& & & & \\
%& & & & \\
%& & & & \\
%& & & & \\
%\hline\hline
%\end{tabular}
%\end{table*}


\bibliography{DGPS-bib}{}
\bibliographystyle{aasjournal}



\end{document}